\documentclass[journal=jctcce,manuscript=article]{achemso}

\usepackage{graphicx} 
\usepackage{wrapfig}
\usepackage{amsmath}

\usepackage{xr}

\makeatletter
\newcommand*{\addFileDependency}[1]{
  \typeout{(#1)}
  \@addtofilelist{#1}
  \IfFileExists{#1}{}{\typeout{No file #1.}}
}
\makeatother

\newcommand*{\myexternaldocument}[1]{%
    \externaldocument{#1}%
    \addFileDependency{#1.tex}%
    \addFileDependency{#1.aux}%
}

\myexternaldocument{SI}

\usepackage{hyperref}
\hypersetup{
    colorlinks=true,
    linkcolor=blue,
    filecolor=magenta,      
    urlcolor=cyan,
    citecolor=blue,
    pdfpagemode=FullScreen,
    }

\title{Improving sampling of binding free energy differences between covalently bound ligands in alternate binding pockets using MT-REXEE}
\author{Anika J. Friedman}
\affiliation{Department of Chemical and Biological Engineering, University of Colorado Boulder, Boulder, CO 80309}
\author{Jerome M. Fox}
\affiliation{Department of Chemical and Biological Engineering, University of Colorado Boulder, Boulder, CO 80309}
\author{Michael R. Shirts}
\affiliation{Department of Chemical and Biological Engineering, University of Colorado Boulder, Boulder, CO 80309}
\email{michael.shirts@colorado.edu}

\begin{document}

\maketitle

\begin{abstract}
    The primary limitation for the application of alchemical free energy methods to a wider variety of complex molecular systems is achieving reasonable sampling. Flexible binding complexes often have high free energy barriers, which require prohibitively long simulations or carefully tuned enhanced sampling methods in order to gather sufficient uncorrelated samples to obtain reliable free energy estimates. An example of such a flexible system is the complex formed between FabB, an elongating $\beta$-ketoacyl-acyl carrier protein (ACP) synthase (KS) from \textit{Escherichia coli}, and ACP, which carries acyl chains of varying lengths. Previous experimental evidence suggests that growing acyl chains can bind to at least two pockets in FabB. With the multiple topology replica exchange of expanded ensemble (MT-REXEE) enhanced sampling approach, we can obtain highly efficient sampling of both pockets by adaptively growing and shrinking the chains in the simulation ensemble, allowing each simulation to visit chain lengths where transitions between the pockets occur. This enables unbiased sampling of alternate configurational states for large complex systems without prior pocket definitions, as collective-variable based enhanced sampling methods would require. Using the new swapping approach gives significantly enhanced sampling even for this simpler problem, as demonstrated by faster convergence of free energy estimates of relative binding affinity between kinetically separated binding pockets. This case study demonstrates the utility of MT-REXEE and its open-source implementation for systems that feature high free energy barriers for a subset of ligands of interest, demonstrating a valuable addition to the existing stable of enhanced sampling methods.
\end{abstract}

\section{Introduction}
Alchemical free energy (FE) calculations have become standard in many computational molecular researchers' workflows, but their application is primarily limited to systems with low conformational flexibility~\cite{yorkModernAlchemicalFree2023, choderaAlchemicalFreeEnergy2011a}. Historically,  FE simulations have commonly supported binding free energy calculations for small-molecule drug discovery~\cite{abelCriticalReviewValidation2017, mueggeRecentAdvancesAlchemical2023}, where they enable drug optimization based on the strength of noncovalent binding to a protein. Covalent drugs represent a new frontier; they offer the potential for increased potency and duration of action~\cite{daltonMedicinalChemistryPerspective2025}, but are not suited for traditional computational methods designed for small molecules. Covalent drug compounds are often larger and more flexible and also have the added challenge of sampling constrained by the covalent bond~\cite{decescoCovalentInhibitorsDesign2017, mueggeRecentAdvancesAlchemical2023, schaeferRecentAdvancesCovalent2023, singhResurgenceCovalentDrugs2011}. In many cases, these systems can remain trapped in metastable states, requiring prohibitively long simulations to escape and properly sample the configurational space. These limitations have prompted the development of numerous enhanced sampling methods, including our own multiple topology replica exchange of expanded ensemble~\cite{friedmanMultipleTopologyReplica2025}.

Though drug discovery is the most common application of free energy methods, covalent ligands are also important in protein and enzyme engineering, where relevant systems are often even larger and more flexible than those encountered in drug discovery efforts~\cite{laserraAlchemicalFreeEnergy2022, papadourakisAlchemicalFreeEnergy2023}. Engineered protein systems often involve multimeric complexes, which include flexible inter-protein interfaces. Fatty acid synthases (FASs) provide abundant examples. These biocatalytic assembly lines carry out the iterative biosynthesis of the fatty acids required for enzyme assembly and energy storage, and have served as powerful platforms for the microbial synthesis of oleochemicals ~\cite{singhProductionBiofuelsOptions2022, pflegerRecentProgressSynthesis2023, peoplesKineticFrameworkModeling2022}. 

$\beta$-ketoacyl-acyl carrier protein synthases (KSs), which catalyze the carbon-carbon bond formation at the heart of fatty acid synthesis, form a covalent intermediate and provide a nice model system for testing FE methods particularly for covalent ligands. Elongation begins when an acyl-ACP associates with a KS, enabling transfer of the acyl cargo on its Ppant arm to the active site cysteine of the KS, producing a covalent acyl-KS adduct and holo-ACP~\cite{mindreboGatingMechanismElongating2020a}. Sampling enhancements are more straightforward to quantify for the transacylation state than for the enzyme-ACP complex, yet a change in the stability of this intermediate state would affect the catalytic activity of the enzyme in a similar manner to perturbing the stability of the ACP complex~\cite{buyachuihanHowAcylCarrier2024, chanCurrentUnderstandingFatty2010}. 
In the covalent intermediate, sampling issues arise primarily from the presence of a large, flexible substrate whose conformational flexibility is restricted by the covalent bonding to the protein and a narrow primary binding pocket. In a separate study, we focused simulations on the enzyme-ACP complex to support a biophysical interpretation of experimental findings, rather than as a verification of our sampling method~\cite{jiangHiddenBindingPocket2026}.

Common enhanced sampling methods have significant disadvantages when applied  for the flexible covalent ligands of KSs. One such set of methods uses an alchemical variable present in free energy calculations, $\lambda$, to enhance sampling; these include Hamiltonian replica exchange~\cite{jiangReducedFreeEnergy2018} and expanded ensemble~\cite{zhangExpandedEnsembleMethods2021, abergDeterminationSolvationFree2004} approaches. These methods require that either the end states or the intermediate states have significantly different free energy barriers in configuration space than the purely physical states. Performing large chemical modifications is a known limitation of these enhanced sampling methods. As such, if we were to perform a single alchemical transformation for a 12-carbon acyl chain, where all particles are alchemically changing at the same time, we would expect slow kinetics for virtually all of the intermediates because of the many intermediately interacting particles present at all stages of the calculation. As a result, these simulations would likely be difficult to converge and produce large error ranges for the free energy estimates. In addition, for KS enzymes, we are interested in the free energy of binding as a function of acyl chain length. Performing a single alchemical transformation from short to long chains (e.g. 4 to 16 carbons) would not give information about the intermediate chain lengths, and performing isolated simulations of smaller transformations (i.e. 4 to 6, 6 to 8, etc.) would not result in sufficiently low FE barriers between alternate bound configurations at long chain lengths to converge sampling. 

Collective variable (CV) dependent methods, including metadynamics~\cite{min_convergence_2007, hsu_alchemical_2023}, umbrella sampling~\cite{you_evaluating_2017, torrie_nonphysical_1977}, and weighted ensemble approaches~\cite{zwierWESTPAInteroperableHighly2015}, represent alternative approaches to enhanced sampling.  However, they require significant prior knowledge about the conformational changes that enable transition between trapped metastable states. For the FAS examples, although we know that longer chains must be able to move when bound, the number of highly correlated degrees of freedom involved makes it challenging to identify relevant CVs that can describe such a transition involving all of the atoms. Intermediate motions along the a non-ideal proposed CV likely sample high-energy configurations, which will decrease sampling efficiency. Even with an approximate path, these CV-dependent methods require prior knowledge of where any alternate binding pockets may be or significant trials and sophisticated, problem-specific data analysis to derive a more complete CV. Given the necessity of this prior knowledge, CV-dependent methods are not well suited for exploratory simulations for which alternate binding pockets may be unknown or poorly defined, which is the case for the FabB-acyl complex.

We previously presented the enhanced sampling method, multiple topology replica exchange of expanded ensemble (MT-REXEE), which specifically targets enhancing sampling in flexible receptor-ligand binding systems~\cite{friedmanMultipleTopologyReplica2025}. In this work, we present improvements made to the MT-REXEE method through the introduction of a novel swapping algorithm, random range swapping, which increases the efficiency of the calculations and improves FE estimate convergence. This method works by enabling trajectory frames other than the final frame in the iteration to be used as potential swappable states in place of the previous solution of simulating  redundant end-states. We use this improved algorithm for MT-REXEE to target the difficult problem of calculating the binding free energy acyl chain length dependence for the KS-acyl complex. Using this problem, we demonstrate the ability of MT-REXEE to significantly enhance conformational sampling in bound systems by coupling free energy simulations together configurationally.

The elongating KS from \textit{Escherichia coli (E. coli)}, FabB~\cite{ruppeKineticRationaleFunctional2020}, provides an excellent test case for demonstrating how MT-REXEE can accelerate sampling for constrained covalent ligands. FabB can generate broad product profiles in the presence of G107M, an engineered mutation that blocks the primary binding pocket (pocket A) and eliminates elongation beyond 8 carbons in the analogous enzyme FabF~\cite{mainsKetosynthaseMutantsEnable2023a}. Crystallographic data shows an alternative binding pocket (pocket B), which allows medium chains to bind when the canonical binding pocket is blocked by G107M~\cite{jiangHiddenBindingPocket2026}. In this paper, we demonstrate the ability of MT-REXEE to sample pockets A and B in both WT and G107M FabB variants, without applying a directed bias towards the pocket. This lack of bias enables the discovery of additional conformational states, such as in this instance a third pocket C sampled a small fraction of the time, and acts as a proof of concept that MT-REXEE can enable the discovery of novel conformational states. With MT-REXEE, not only do we access both pockets within the same simulation, but we can also quantify the FE penalty of the G107M mutation in pocket A and the relative stability between pockets A and B.

Sampling of both distinct binding pockets in this system is not possible using standard MD or independent expanded ensemble simulations within reasonable simulation lengths ($<$500 ns) for substrates longer than 4 carbons. Nonetheless, we can leverage the increased conformational flexibility of the short-chain substrates through these conformational swaps to increase sampling in long-chain simulations. We present this method as a suitable approach when only a subset of ligands have a high configurational free energy barrier to overcome. We also demonstrate how MT-REXEE simulations can produce both configurational and alchemical free energy values for all evaluated end points and between two or more distinct binding pockets.

\section{Theory: Separating Alchemical and Conformational Free Energies}
\begin{figure}[H]
    \centering
    \includegraphics[width=0.7\linewidth]{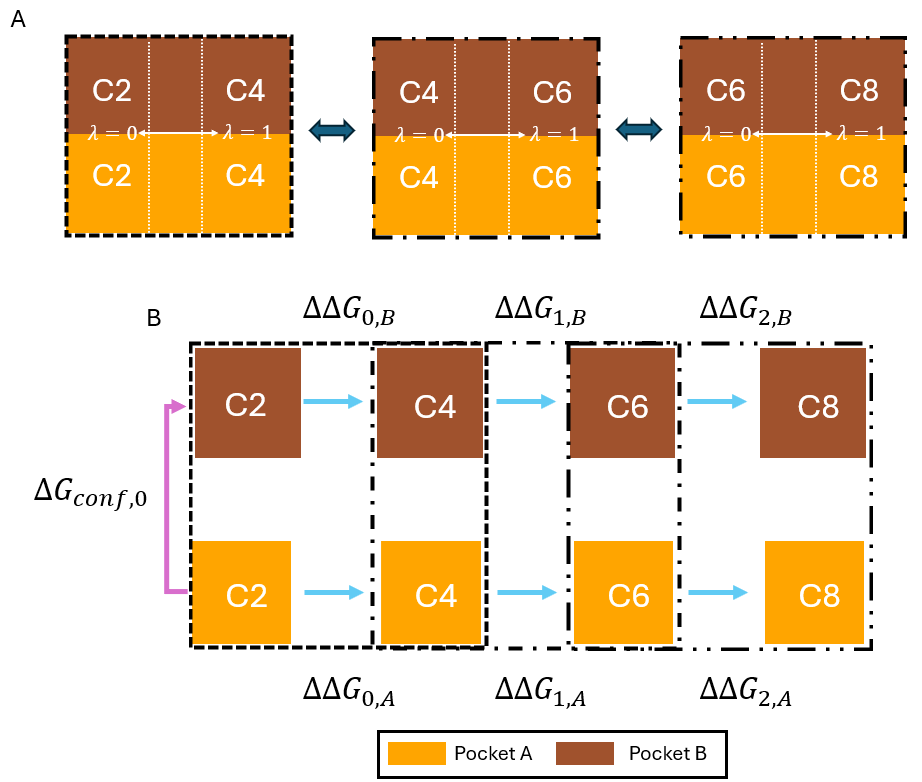}
    \begin{singlespace}
    \caption{\label{fig:total_transform}(A) A set of MT-REXEE simulations that perform transformations of the acyl substrate from C2 to C4, C4 to C6, and C6 to C8. Each simulation, which visits a range of alchemical states $\lambda$ in [0,1], has configurations that should (in the limit of sufficient sampling) visit both pocket A (orange) and pocket B (brown). The dashed boxes indicate states in each individual simulated transformation, and the dark blue arrows show where configurational swaps can occur, at the alchemical end states. (B) When we separate the transformations by end state and configurational state, we can estimate each of the FE differences we compute from a MT-REXEE simulation. We compute the alchemical FE difference for each transformation independently in each pocket (light blue) as well as the configurational FE difference at at least one end state (pink). This enables us to compute the relative FE difference between any combination of alchemical end state and configurational state.}
    \end{singlespace}
\end{figure}

One application of the MT-REXEE method is to reconstruct the binding FE dependence on both ligand identity and bound conformation for systems with two or more distinct binding locations. Here, we apply MT-REXEE to the covalent intermediate of FabB, as this system features a wide variety of ligand substrates (acyl chains of 4-16 carbons) and is expected from crystallographic evidence to sample two distinct acyl chain binding locations (A and B)~\cite{jiangHiddenBindingPocket2026}. It is important to note that the expressions below apply for computing the free energy differences between two states regardless of how many possible conformational states there are for the complex (e.g., we ignore the transient binding pocket C). For this system, each MT-REXEE simulation performs one alchemical transformation and can access two or more configurational states (Figure \ref{fig:total_transform}A). MT-REXEE enables conformational exchange between each transformation, enabling more comprehensive sampling of the two or more conformational states. From the set of MT-REXEE simulations, we obtain the dependence of binding FE on both the substrate and the binding conformation (Figure \ref{fig:total_transform}B).

First, we demonstrate that we can partition the conformational space to calculate the alchemical free energy difference between two end-points independently for each configurational state. This calculation will provide the values for $\Delta\Delta G_{n,A}$ and $\Delta\Delta G_{n,B}$ (light blue arrows) in figure \ref{fig:total_transform}B, where $n$ indicates the transformation. We start with the alchemical Gibbs free energy difference given by equation \ref{eq:core_FE} where $\beta = \frac{1}{k_B T}$, $k_B$ is the Boltzmann factor, $T$ is the temperature, $U_i$ and $U_j$ are the potential energies as a function of $\vec{q}$ the coordinates and momenta for two alchemical states and $\Gamma$ denotes the phase space volumes of $\vec{q}$ over which we sample which can generally be assumed to be equivalent for both states ($\Gamma_i = \Gamma_j$).
\begin{equation}\label{eq:core_FE}
    \Delta G_{ij} = -\frac{1}{\beta}\ln{\frac{Q_j}{Q_i}} = -\frac{1}{\beta}\ln{\frac{\int_{\Gamma}e^{-\beta U_j(\vec{q})}d\vec{q}}{\int_{\Gamma}e^{-\beta U_i(\vec{q})}d\vec{q}}}
\end{equation}
The total phase space can be decomposed into the space consistent with sets of conformations in pocket A and B, such that $\Gamma_i = \Gamma_{i,A} + \Gamma_{i,B}$. Here we are interested not in the overall relative free energy difference between the two alchemical states ($i$ and $j$) but rather the relative free energy difference between the two alchemical states within the restricted phase space consistent with configurational state A and thus equation \ref{eq:core_FE} becomes \ref{eq:new_core}. $\Delta G_{ij, A}$ is the free energy difference of performing the alchemical transformation from state i to j while in pocket A, and $\Delta G_{ij, B}$ would give the same transformation but in pocket B. 
\begin{equation}\label{eq:new_core}
    \Delta G_{ij, A} = -\frac{1}{\beta}\ln{\frac{Q_{j, A}}{Q_{i, A}}} = -\frac{1}{\beta}\ln{\frac{\int_{\Gamma_{A}}e^{-\beta U_{j, A}(\vec{q})}d\vec{q}}{\int_{\Gamma_{A}}e^{-\beta U_{i, A}(\vec{q})}d\vec{q}}}
\end{equation}

The alchemical free energy difference within conformational space $A$ and $B$ can be calculated by dividing the trajectory frame by frame into two non-continuous trajectories corresponding to spaces $A$ and $B$ based on the configurational state of each frame. We can then perform two separate MBAR calculations (Figure \ref{fig:FabB_FE}) with the samples from each trajectory. The two partition functions are additive, but the two alchemical free energies extracted from each transformation are not additive, which we expand upon in section \ref{SI:fe_add}.

There may be concern that the dimensionality of the conformational space defining A and B will vary between transformations, as each transformation is performed with different numbers of dummy particles. At this point, it is important to note that this method is being used purely for differences in free energies between two environments. In these calculations, an equivalent transformation between states $i$ and $j$, is performed in solvent in addition to the transformation in complex with the protein. Since the calculation of $\Delta\Delta G$ is independent for each transformation and we maintain only three bonded interactions between the dummy and "real" atoms during the transformation, Boresch shows that the dummy atom contributions cancel, assuming equal dummy atom treatment in both solvent and protein complex~\cite{boreschRoleBondedEnergy2002}. We could rectify the issue of dimensionality differences in partition space through the addition of dummy atoms to all transformations, such that all atoms necessary for all transformations are present, but many remain in the dummy state for both end-points of the FE transformation. 

In this current application, the distinction between pocket A and B exists solely in the protein complex conformation and thus in solvent $\Delta G_{ij, A, solv} = \Delta G_{ij, B, solv}$ but this is not necessarily the case in general, so the two will be left as independent variables in the derivation for the purpose of generalization. Thus, the relative binding free energy between alchemical states $i$ and $j$ ($\Delta\Delta G_{ij, A}$) can be given by equation~\ref{eq:ddG_A} (see Figure \ref{SI:cycle} for the full thermodynamic cycle). For simplicity, when dealing with multiple subsequent transformations, we instead use the notation $\Delta\Delta G_{n, A}$ where n denotes the transformations in order from the base transformation.
\begin{equation}\label{eq:ddG_A}
    \Delta\Delta G_{n, A} = \Delta G_{n, A, complx} - \Delta G_{n, A, solv}
\end{equation}

Finally, we need to compute a configurational FE difference between pockets A and B, i.e., $\Delta G_{conf, 0}$ (pink arrow) in Figure \ref{fig:total_transform}B. In standard MD simulations, which sample from the ``native'' Boltzmann distribution with no conformational biases, the relative probability of sampling within the conformational space of pocket A versus pocket B is given by the integrated Boltzmann factor over the phase spaces volumes corresponding to the configurations; the configurations where to the chain occupies pocket A versus the set of configurations where the chain occupies in pocket B.  Assuming proper Boltzmann sampling in the MD simulations, then this ratio of configurational partition functions becomes simply the ratio of time (i.e. number of frames) spent in each configurational state. This is shown in Equation~\ref{eq:boltzmann_volumes} where $p_A$ and $p_B$ are the probability of sampling within the conformational space of pocket A and pocket B respectively and $\Gamma_A$ and $\Gamma_B$ are the sets of configurations corresponding to the chain being in pocket A and B respectively. 
\begin{eqnarray}
    \frac{p_A}{p_B} &=& \frac{Q_A}{Q_B} =\frac{\int_{\Gamma_A} e^{-\beta U(\vec{x})} d\vec{x}}{\int_{\Gamma_B} e^{-\beta U(\vec{x})} d\vec{x}}\label{eq:boltzmann_volumes}\\
    \Delta G_{conf, AB} &=& -k_B T \ln \frac{p_A}{p_B}\label{eq:conf_fe}
\end{eqnarray}

In the specific application in this paper, we use our expanded ensemble (EE) alchemical free energy simulations to compute the configurational free energy $\Delta G_{conf, AB}$ rather than using standard MD. Though it may be possible to incorporate the non-physical alchemical intermediate states into the calculation of the configurational free energy difference, for simplicity we restrict the calculation to physical end states. We extract the subset of the sampled configurations from the physical end states ($\lambda=0$ and $\lambda=1$), and we can apply equation \ref{eq:boltzmann_volumes} to determine the configurational free energy difference. We require only one configurational FE difference between states for one of the ligand identities, as we can use the thermodynamic cycle and the alchemical differences to find the configurational difference for any other chain length. The choice of state to determine the configurational free energy difference should be the state for which the kinetics of exchange between configurational states is the fastest. In the current study, this is the C2 acyl chain end state. This state had statistically equivalent pocket sampling to that of the independent MD simulations (see Figure \ref{fig:fabb_summary}B), and the uncertainty in this step is less than the uncertainty in the alchemical process~\cite{friedmanMultipleTopologyReplica2025}. Thus we can use the relative probability of the C2 chain sampling either pocket A or B with equation \ref{eq:conf_fe} to compute the configuration FE difference $\Delta G_{conf, 0}$.

This configurational FE difference ($\Delta G_{0, conf}$) along with the alchemical free energy differences in each transformation ($\Delta\Delta G_{n,A}$ and $\Delta\Delta G_{n,B}$) can then be used to compute the FE difference between any end state in any configurational state. Thus, the configurational free energy difference for an end point after $x$ transformations is given by \ref{eq:dg_conf_gen} (Figure \ref{SI:total_FE}). This relationship will change if the choice of flexible conformational state used to compute the configurational FE difference is different, but we can use any sum of paths which connect the two relevant end states.
\begin{equation}\label{eq:dg_conf_gen}
    \Delta G_{x,conf} = \Delta G_{0, conf} + \sum_{n=0}^{x}\Delta\Delta G_{n,A} - \Delta\Delta G_{n,B}
\end{equation}

\section{Methods}
\subsection{System Set-up}
We used two systems to test our approaches: (i) the complex formed between mouse major urinary protein 1 (MUP1) and thiazole-based pheromones which we previously studied with MT-REXEE\cite{friedmanMultipleTopologyReplica2025} and (ii) the FabB-acyl complexes described in the introduction. The MUP1-thiazole complex contains a cyclic ligand bound to a rigid $\beta$-barrel protein domain for which convergence of free energy estimates requires easily accessible computational resources~\cite{robo_fast_2023}. This set of simulation provide verification of the newly implemented random range swapping method and demonstrate the capability of MT-REXEE to enhance sampling even in simple binding free energy calculations. The FabB-acyl complex is an ideal example to demonstrate the conformational sampling enhancement of the MT-REXEE approach. This system features high free energy barriers for acyl chain rearrangement, which create trapped conformational states for which the transitions are prohibitively difficult to sample with existing methods. We derived the MUP1 complexes from the crystal structure this protein bound to 2-sec-butyl-4,5-dihydrothiazole (PDB 1I06). We derived the FabB-acyl complexes from the crystal structure of FabB covalently bound to a C12 acyl chain (PDB 1EK4). For FabB, we used MODELLER to apply the G107M mutation and we used Avogadro to shrink or elongate the acyl chain to generate C4-16 initial configurations. All small molecules were parameterized using GAFF2~\cite{heFastHighqualityCharge2020} and the protein was parameterized using the Amber \textit{ff14SB} force field\cite{maierFf14SBImprovingAccuracy2015a}. The relative binding free energy topologies for all systems were created using PMX~\cite{gapsysPmxWebserverUser2017a} followed by post-processing to remove all dihedral energies between dummy and real atoms within the hybrid topologies. The treatment of dummy atoms in this study follows the practices presented in earlier works to minimize any effects on the partition function that are not canceled out by alchemical transformations in different environments~\cite{fleckDummyAtomsAlchemical2021a, meyBestPracticesAlchemical2020a}. The unit cell boxes for all simulations conducted in solvent were constructed to maintain a minimum distance of 1 nm between the molecule and the periodic boundary and then filled with TIP3P~\cite{jorgensenComparisonSimplePotential1983} solvent. All simulations were run in a modified version of GROMACS 2022.5~\cite{bauerGROMACS20225Source2023}, which can be found as a branch on the official GROMACS \href{https://gitlab.com/gromacs/gromacs/-/tree/add_wl_histogram_input_2022_5?ref_type=heads}{GitLab repository}. This modification allows the acceptance of the MDP parameter \texttt{init\_histogram\_counts}, which is crucial for performing Wang-Landau weight equilibration across iterative trajectories. Importantly, all code changes in this branch have been merged into the main code base as of GROMACS 2025.0. The MT-REXEE implementation used was version 1.1.0 which can be found within the \textit{ensemble\_md} GitHub package~\href{https://github.com/shirtsgroup/ensemble_md}{github.com/shirtsgroup/ensemble\_md}.

We used the same procedure for equilibrating all systems. This procedure consisted of an initial energy minimization to a threshold of 50 kJ/mol$\cdot$nm followed by a 100 ps NVT equilibration using the Bussi-Parrinello thermostat~\cite{bussiCanonicalSamplingVelocity2007a} maintaining a temperature of 300 K. A 100 ps NPT equilibration was also performed with the stochastic cell rescaling barostat~\cite{bernettiPressureControlUsing2020a} for all calculations performed in solvent to maintain a pressure of 1 atm. All simulations were performed with a 2 fs time step. Additional simulation set-up information can be found for each simulation in the project GitHub repository~\href{https://github.com/shirtsgroup/MT-REXEE}{github.com/shirtsgroup/MT-REXEE}.

We initiated the FabB-acyl complex with the acyl chain in two different, non-overlapping pockets, which are referred to as pocket A and pocket B (Figure \ref{fig:fabb_summary}A). An initial standard MD simulation was performed for the FabB-acyl complex in pocket A with 50 ns at $\lambda=0$. The purpose of this simulation was to extract decorrelated frames from the MD trajectory to use as initial configurations for our MT-REXEE simulations. A duration of 50 ns was selected to increase the probability of sampling alternate metastable states. The mean transition time between pockets A and B for the 2 carbon chains was approximately $11\pm 5$ ns, which gives us an approximation for the timescale of significant acyl chain rearrangements. We sought to extract three frames separated by 11-16 ns of MD simulation, requiring a total of 48 ns, which was rounded to 50 ns for simplicity. From this simulation, three random frames were selected in the first, second, and last third of the trajectory as the initial configuration for each of the three replicas. These random frames were verified to have an acyl chain RMSD of at least 5\AA~to generate substantial initial conformational diversity. We then run MT-REXEE simulations for the FabB-acyl complex in pocket A. As shown below, these simulations sample both pocket A and pocket B. We extract one frame in which each complex samples pocket B from these MT-REXEE simulations. We then ran 50 ns of standard MD on each of these pocket B configurations at $\lambda=0$ and again extracted a random frame from the first, second, and last third of the MD trajectories as the initial configurations. 

\subsection{Free Energy Protocol}
We performed both EE and MT-REXEE simulations in triplicate for each system. The EE settings used were identical for both the MT-REXEE and independent EE simulations with a Wang-Landau scale of 0.82 and a Wang-Landau ratio of 0.85 for the equilibration for the FabB system, and 0.8 was used for both settings to calculate the free energy of ligand binding to MUP1. These parameters were tuned to ensure relatively even sampling between $\lambda$ intermediate states following weight equilibration. Slower weight equilibration through the increasing of Wang-Landau scale and ratio parameters extends equilibration times but produces more even $\lambda$ state sampling once equilibrated weights are obtained. The weight equilibration phase was ended when the Wang-Landau $\delta$ was less than 0.001 for both systems. The MUP1 ligands were run for 30 ns and 50 ns after equilibration for the solvent and protein complex simulations, respectively. For the FabB-acyl complexes, which represent an intermediate state preceded by FabB bound to acyl-ACP, we defined the``solvent'' (i.e., pre-complex) simulation state as acyl-ACP in solvent. The FabB systems were run for 50 ns and 100 ns for the solvent and enzyme-bound states, respectively. All free energy estimates were calculated using the MBAR implementation in alchemlyb~\cite{wuAlchemlybSimpleAlchemistry2024}. 

We also ensured that the overlap between adjacent $\lambda$ states was sufficient for all transformations both in solvent (Figure \ref{SI:ACP_solv_overlap}) and in complex (Figure \ref{SI:FabB_cmplx_overlap}). We used the same $\lambda$ state spacing for both the EE and MT-REXEE simulations, as well as the same initial configurations for all replicas.

The primary feature added to the MT-REXEE package in this study is the use of a new swapping criterion between simulations. In the previously developed exhaustive approach, swaps are only attempted for the last frame of each iteration. This new swapping criterion is named random range (RR), and rather than only allowing swaps between trajectories in the last frame of an iteration, we instead allow swaps at random compatible frames (i.e. endpoints of two neighboring simulations that represent, up to the presence of different dummy atoms, the same physical system) in a defined portion of the trajectory. We used the last 50\% of the trajectory to result in longer continuous sampling within each trajectory, decreasing the structural correlation to the previous iteration. Regardless of the location of the trajectory used for swapping, all data from each iteration is used to compute free energy estimates and any additional analysis; only trajectory swapping and continuation use the randomly sampled exchange-compatible frame. For the MUP1 system, we used redundant end states (5 intermediate $\lambda$ states as well as 5 states each at $\lambda$=0 and $\lambda$=1) in combination with the exhaustive swapping method as we did in our previous MT-REXEE paper~\cite{friedmanMultipleTopologyReplica2025} to compare to the random range swapping approach without redundant end states. For the FabB-acyl systems, we ran only MT-REXEE simulations with the RR method and with single $\lambda$ end-states.

\subsection{Configurational Analysis}\label{sect:methods_config}
X-ray crystal structures of the FabB--acyl-ACP complex show two distinct binding pockets, for the acyl chain. We speculated that these two pockets, termed A and B, would also be occupied in the covalent intermediate. It is important to note that though we had a similar complex crystal structure to inform our definitions of the binding pockets, these pocket definitions and classifications do not affect the MT-REXEE simulations themselves and are applicable only to the analysis of the simulations. To develop an unbiased classifier of pocket occupancy, we used the MT-REXEE simulations with the acyl chain initialized in the canonical binding pocket (A) and visually labeled a subset of trajectory frames (12--15) for all chain lengths as being in either pocket A or B. We determined that the distance between residue THR300 and carbon 4 in the acyl chain could be used to differentiate the two pockets (Figure \ref{SI:pocket_dist}A). We evaluated several distance thresholds between pairs of atoms in the acyl chain and FabB as well as acyl internal dihedrals as metrics for pocket determination, but the distance between C4 and THR300 had the least overlap between pockets A and B. We plotted the distance distribution of these residues in EE simulations, which were visually verified to have all frames within either pocket A or B, as well as for frames from our MT-REXEE simulations, which sample both pockets, to determine the threshold of 0.85 nm (Figure \ref{SI:pocket_dist}B). Several conformations for longer chains (8 or more carbons) were identified as belonging in pocket A or B by this criterion, but did not match existing known structures for the pockets. Thus, we also applied a chain compaction criterion that the distance between carbons 2 and 8 should be shorter than 0.6 nm in pocket B and longer than 0.75 nm in pocket A, limits based on crystal structures. Any trajectory frames in which this distance to THR300 is greater than 0.85 nm and the compaction distance is greater than 0.75 nm are considered in pocket A. Any frames with a distance to THR300 less than 0.85 nm and a compaction distance of less than 0.6 nm are considered in pocket B. Any frames for which neither pocket criteria are satisfied were discarded from free energy analysis, as they are in neither pocket.  For the C2 and C4 complexes, these discarded frames were 1-4\% of the trajectories, and for all other chain lengths, the discarded frames were $<$ 0.5\% of the trajectories.

\section{Results}
\subsection{Random range swapping improves round-trip time and statistical convergence compared to previous exchange methods}
\begin{figure}[H]
    \centering
    \includegraphics[width=0.9\linewidth]{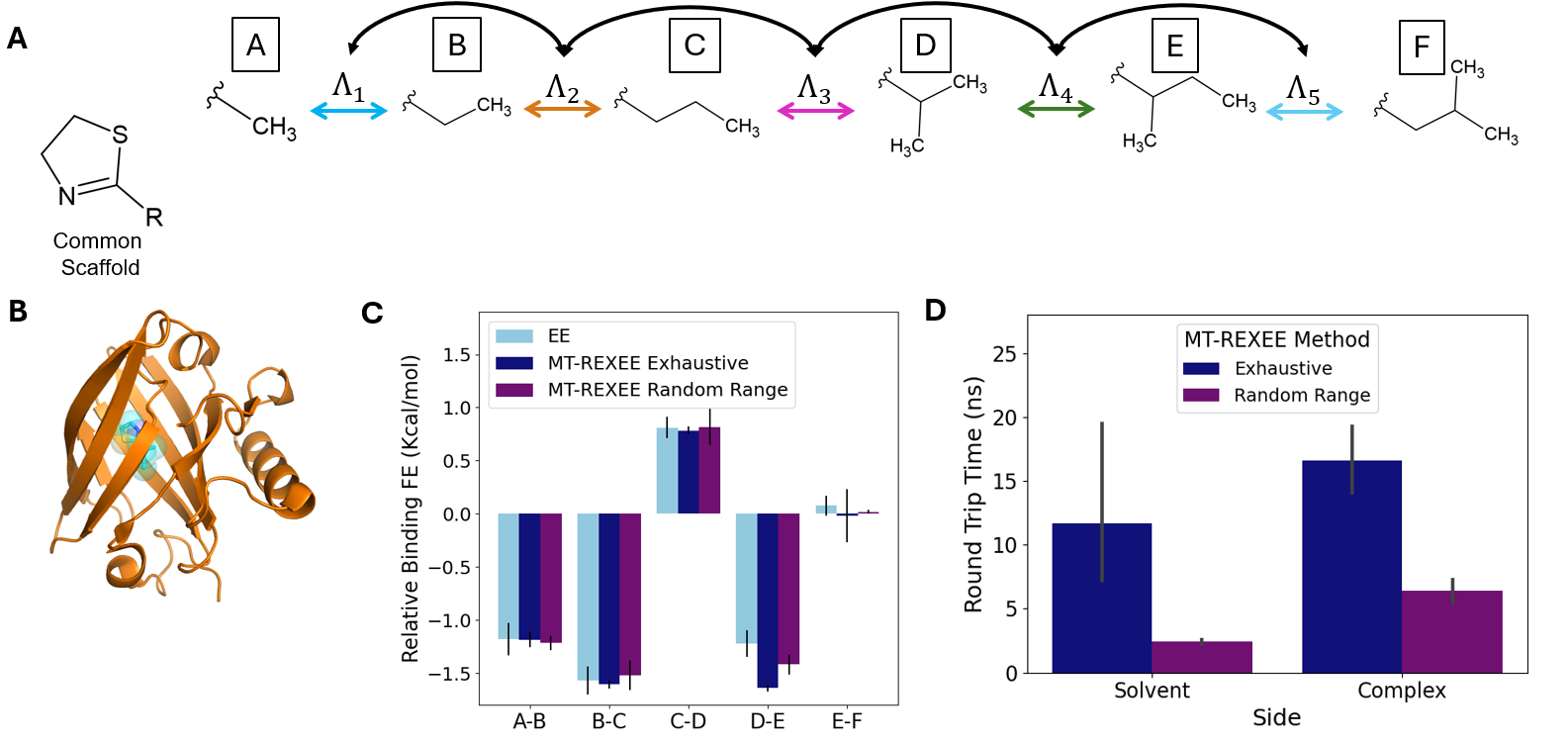}
    \begin{singlespace}
    \caption{\label{fig:mup1}Random range swapping yields statistically equivalent free energies while significantly decreasing the round-trip time through alchemical states. (A) A diagram of the sequence of ligand transformations used in binding free energy estimates with the protein MUP1 to validate the random range swapping scheme. The curved arrows show where MT-REXEE swaps occur. (B) The structure of MUP1 in complex with ligand A. (C) We compare the binding free energy estimates using EE, MT-REXEE with the exhaustive swapping scheme, and MT-REXEE with the RR swapping scheme. We observe no significant differences between the free energy estimates for all three methods as the values are all within statistical error of one another. (D) We observe a significant (p$<$0.001) decrease in the round-trip times from 10-20 ns to 3-5 ns for both solvent and MUP1 complex simulations using the RR swapping scheme compared to the previous exhaustive scheme.}
    \end{singlespace}
\end{figure}

In this section, we compare the results for running binding free energy calculations for a set of small molecule ligands with the protein MUP1 using both the previously introduced exhaustive method and the random range (RR) method (Figure \ref{fig:mup1}A--B). We ran this comparison to determine any changes in performance and ensure that this new algorithm for selecting swappable states does not introduce additional error. The RR method no longer requires redundant end states. This eliminates the increase in Wang-Landau weight equilibration times observed in our earlier work~\cite{friedmanMultipleTopologyReplica2025}. The result of this improvement is that MT-REXEE with the RR exchange method (MT-REXEE RR) has equivalent or lower total Wang-Landau weight equilibration times compared to independent EE simulations (Figure \ref{SI:mup1_equil}). We note that weight equilibration times are calculated using the full iteration in which the weights reach the equilibration criteria, even if conformational exchanges occur at a frame earlier than the final frame of the iteration. Only the E-F transformation had a potentially statistically significant difference between the weight equilibration times using EE and MT-REXEE RR ($p=0.083$). We hypothesize that the decrease in weight equilibration times for the E-F transformation is likely due to the improved conformational space sampling using RR. This transformation is the most chemically distinct of all the MUP1 ligands evaluated in this study, and for this reason, it benefits most from the increased conformational sampling.

The binding free energy estimates are statistically equivalent for EE and MT-REXEE with both exhaustive and RR swapping methods, but various convergence metrics indicate that less simulation time is necessary for equal accuracy with RR (Figure \ref{fig:mup1}C). We measured the convergence of free energy estimates both along the length of individual trajectories as well as convergence between trajectories initiated with significantly different initial configurations. Importantly, we observe that for both solvent and, even more significantly for the protein complex simulations, the free energy estimates converge significantly more quickly, resulting in closer estimates between the estimates for the first and last half of the trajectories for MT-REXEE RR compared to both EE and the MT-REXEE simulations using the exhaustive method (Figure \ref{SI:mup1_converge}A--B). We also notice a significant decrease in the standard error of the estimate between replicates of between 40 and 160\% compared to EE (Figure \ref{SI:mup1_converge}C). These independent replicates were initiated with the same initial configurations extracted from standard MD trajectories separated by a minimum of 20 ns. The exchanges between distinct alchemical transformations in the MT-REXEE method enable increased sampling of the conformational space, which produces faster convergence. The standard error is also reduced by between 16-90\% when using the exhaustive swapping method with redundant end states compared to EE. This is likely because the increase in conformational sampling is balanced by the decreased sampling of intermediate $\lambda$ states due to the addition of redundant end states. This results in decreased conformational sampling than MT-REXEE RR, but still faster conformational sampling than EE. This is consistent with the observed significant decrease in the round-trip times for the MT-REXEE RR method compared to the exhaustive swapping method (Figure~\ref{fig:mup1}D). We expect the over 4-fold decrease in round-trip time is primarily responsible for the faster increased conformational sampling in MT-REXEE RR compared to the MT-REXEE exhaustive method. Though MUP1 is a simple test system, it provides additional verification that, like the previous exhaustive swapping method, we are not introducing additional systematic error into the calculations while also showing enhanced sampling in even the simplest systems.

\subsection{Enhanced Sampling of FabB with MT-REXEE}
\begin{figure}[H]
    \centering
    \includegraphics[width=0.7\linewidth]{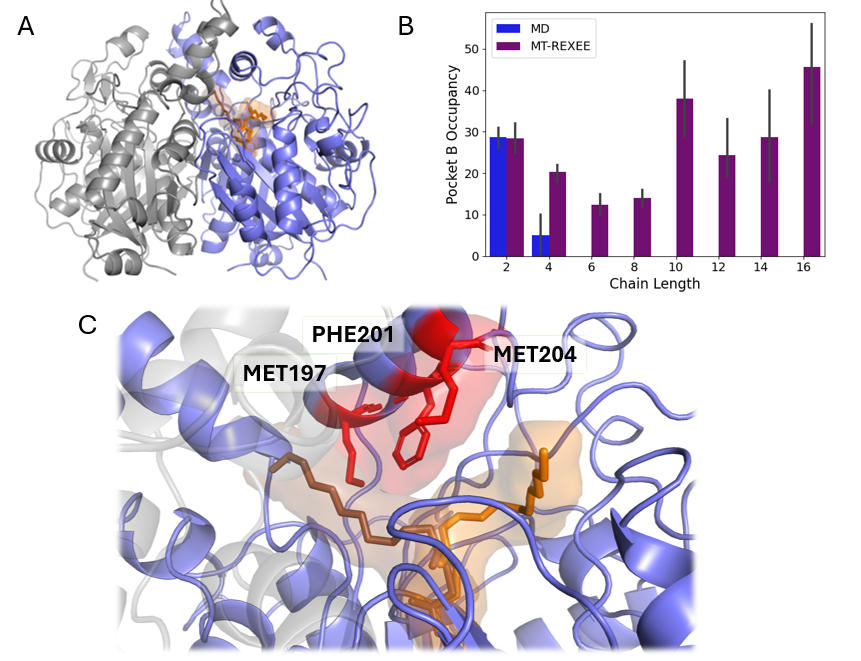}
    \begin{singlespace}
    \caption{(A) The binding pocket of the FabB--acyl covalent intermediate is shown with a 12 carbon acyl substrate in both pocket A (orange) and pocket B (brown) in the context of the overall protein. The structures shown are the centroids from MT-REXEE simulations of the C10-12 transformation for $\lambda=1$. (B)  Occupancy of pocket B as a function of substrate chain length using both standard MD simulations and MT-REXEE enhanced sampling as described in section \ref{sect:methods_config}. Standard MD is not able to access pocket B within 500 ns for any substrate longer than 4 carbons, but we readily see sampling in pocket B for all chain lengths within only 100 ns of MT-REXEE simulation. (C) A close-up image of the binding pocket in (A) highlighting several residues on both the catalytically active monomer and adjacent monomer that restrict the acyl chain's conformational flexibility and prevent interconversion during MD simulations of at least 500 ns.}
    \label{fig:fabb_summary}
    \end{singlespace}
\end{figure}

We also apply the MT-REXEE method to an example system in which there is a known sampling limitation. Prior evidence suggests that when FabB binds acyl-ACP substrates, the acyl chain samples two distinct binding tunnels, pockets A and pocket B, in addition to a transient binding site, pocket C, where the acyl chain is parallel to the Ppant arm--primarily in the enzyme-ACP complex. In this work we focus on pockets A and B, but we expand upon the presence and hypothesized reason for pocket C in a separate work~\cite{jiangHiddenBindingPocket2026}. Here we examine the FabB-acyl intermediate--that is, the complex formed after transfer of the acyl chain from the phosphopantetheine of ACP to the catalytic CYS163 of FabB (Figure \ref{fig:fabb_summary}A). Sampling the FabB-acyl complex is simpler than sampling the FabB--acyl-ACP complex, as the motions of ACP produce additional sampling issues; the FabB--ACP interface is large and flexible. Using the FabB-acyl complex allows us to focus on only the motions of the acyl chain. Additionally, the FabB-acyl complex precedes the condensation reaction with malonyl-ACP, which extends the acyl chain by two carbons, and is a biochemically important state~\cite{buyachuihanHowAcylCarrier2024}. 

Previous mutational data of FabB suggest that pocket B remains accessible in the covalent intermediate, as mutants that block binding to pocket A are still capable of elongating acyl chains up to 16 carbons in length~\cite{mainsKetosynthaseMutantsEnable2023a}. The transition between pockets A and B is too slow to sample with direct simulation using standard MD with reasonable simulation lengths (500 ns; Figure \ref{fig:fabb_summary}B). Therefore, this study uses MT-REXEE with random range swapping to evaluate the relative stability of 2, 4, 6, 8, 10, 12, 14, and 16 carbon length acyl chain substrates covalently bound to FabB. 
We simulated FabB as a dimer with only one substrate attached. This approach is motivated by previous work on multimers of FAS enzymes, where binding of multiple substrates can yield negative cooperativity between monomers~\cite{marcellaStructureHighAffinity2017, priceStructureOfBeta2001}.

We performed the 500 ns standard MD simulations for FabB-acyl complexes with acyl substrates from 2--14 carbons in length, seven simulations in all. All simulations were initiated with the same initial configuration as the corresponding MT-REXEE simulation in pocket A, with the removal of dummy atoms required for MT-REXEE. We ran three replicates with distinct initial configurations for all chain lengths (see Methods). The corresponding MT-REXEE simulations were run for only 100 ns (in parallel) for each transformation of chain length. We then analyzed the trajectories to determine the occupancy of pocket A, B, or neither for each chain length substrate. For MT-REXEE simulations, we specifically compared the pocket occupancy for the end states and not for intermediate $\lambda$ states, as these are irrelevant to the physical occupancies. 

MT-REXEE and standard MD show statistically equivalent sampling of pocket B only for the 2 carbon acyl chain substrate. For all longer substrates, standard MD shows reduced sampling of pocket B, likely because of the high free energy barrier for longer substrates to transition to this pocket (Figure \ref{fig:fabb_summary}B). In contrast, MT-REXEE enables significant sampling in pocket B for all chain lengths studied while utilizing only 20\% of the trajectory length as the standard MD simulations. 

We hypothesize that the inability of standard MD to sample both pockets results from stabilizing interactions between the protein and acyl chain that are present for chains longer than 4 carbons (Figure~\ref{fig:fabb_summary}C). For both MD and MT-REXEE simulations, the 2-carbon acyl chains sample pockets A and B with relatively high frequency. The presence of the conformational swaps between the 2 carbon state and longer chain substrates allows for conformational sampling of both pockets at all chain lengths. MT-REXEE enables simulation trajectories to sample pocket B when simulations are initiated in pocket A, thus requiring no prior knowledge of the presence or location of the alternate pocket. In contrast, existing knowledge of both the location of pocket B and a reasonable transition path would be necessary for employing a configurational variable dependent enhanced sampling method. The lack of directly applied configurational bias along a known CV, such as that used in metadynamics, also provides increased confidence that these alternate conformations are not artifacts of the enhanced sampling method itself. 

In MT-REXEE simulations, switching between pockets is primarily limited by the round-trip time between alchemical simulations of different chain lengths, as we can only switch between pockets in simulations of the shortest chain. Additionally, we want to emphasize that the success of the MT-REXEE method is heavily correlated to the suitability of the applied equilibrated EE weights. An increase in round-trip time correlates with an increased imbalance between $\lambda$ state sampling within the simulations (Figure \ref{SI:wl}). The imbalance in $\lambda$ state sampling can be minimized by adjusting Wang-Landau equilibration parameters. Optimization of these parameters is system-specific. Further optimization of these parameters would likely further minimize this effect, but would also increase weight equilibration times. This necessitates a balance between the efficiency of weight equilibration and the accuracy of equilibrated weights. 

\subsection{Configurational and Alchemical Free Energy}
This sampling of both pockets during the same simulation allows us to achieve two goals:  Estimating the relative free energy as a function of chain length independently in each pocket and estimating the configurational free energy difference for a given chain length substrate in pocket A versus pocket B. The solvent reference state for all alchemical binding free energies is the acyl chain attached to ACP in solvent as described in the Methods section (Figure \ref{SI:ACP_conf}).  

\begin{figure}[H]
    \centering
    \includegraphics[width=0.8\linewidth]{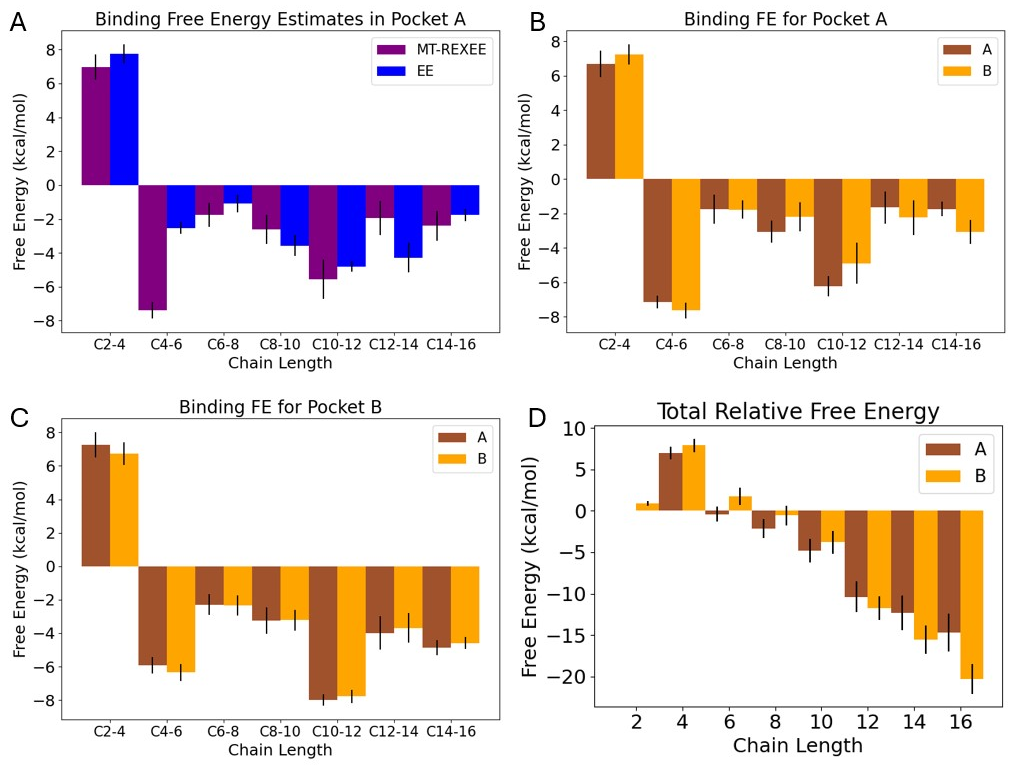}
    \begin{singlespace}
    \caption{The convergence of binding free energy simulations initiated in pockets A and B demonstrates MT-REXEE can increase sampling of the conformational space. (A) Comparison of estimates from individual independent EE calculations with estimates of the same transformations in MT-REXEE simulations. MT-REXEE's significantly improved sampling leads to moderate differences between the two methods, with MT-REXEE expected to be more accurate. Estimates for binding free energy using MT-REXEE with the initial acyl chain configurations in either pocket A (brown) or pocket B (orange) for the calculation of the RFE in pocket A (B) and pocket B (C), showing independence of starting configuration. (D) Cumulative relative binding free energy can be obtained for all chain lengths relative to a C2 chain for both pockets using simulations initiated in both pockets A and B (Figure \ref{SI:total_FE}). Thus, from a single set of MT-REXEE simulations, we can obtain the full chain length and pocket dependence on binding FE. For all figures, we plot the mean of n=3 (n=6 for D) replicas and error bars represent the standard error of the mean.}
    \label{fig:FabB_FE}
    \end{singlespace}
\end{figure}

The EE simulations were initiated with the same chain conformations as the MT-REXEE simulations in pocket A. Since the EE transformations initialized in pocket A are only able to sample pocket A, we can only make comparisons to the alchemical free energy transformations in pocket A. We observe a significant difference between the free energy estimate for the C4 to C6, C6 to C8, and C12 to C14 transformations when using MT-REXEE compared to EE (Figure \ref{fig:FabB_FE}A). It is not unexpected that some transformations would produce significantly different FE estimates as even within a given pocket the configurational space sampled with each of the two methods is expected to vary greatly. This is less likely to affect the known flexible transformations like C2 to C4, which would be expected to readily sample the conformational space in both methods. The degree of sampling with a given binding pocket is also less likely to affect the longest substrates which have fewer accessible conformations that fit the full acyl chain. Overall, these results are consistent with the significantly different conformational sampling expected with MT-REXEE compared to EE. We would worry about MT-REXEE introducing additional error into the calculations producing these differences; however, the fact that the C2 to C4 transformation, the one transformation that we would expect EE alone to sample effectively, is statistically identical between the two methods suggests no errors are introduced.

We also performed MT-REXEE simulations initiated in pocket B to assess the convergence of the free energy estimates between the pockets, when the simulations were started in pocket A versus started in pocket B. After 100 ns of MT-REXEE simulation, the estimates were within statistical error regardless of whether they started in pocket A (Figure \ref{fig:FabB_FE}B) or pocket B (Figure \ref{fig:FabB_FE}C), with the possible exception of the C14 to C16 transformation initialized in pocket A. A single simulation of this transformation, when initiated in pocket B, showed significantly less sampling in pocket A than the other two replicas, and contributes significantly to this deviation. If simulations were extended additional metastable states may emerge, so greater sampling difficulty for very long chains is expected. 

Having established convergence between the two sets of initial configurations, all future presented data will use both sets of simulations regardless of initialization in pocket A or B. From these combined simulations, we compute the cumulative free energy curve with all estimates relative to the binding free energy of a C2 acyl chain in pocket A (Figure \ref{fig:FabB_FE}D). 

Our results indicate that for acyl chains of 10 carbons or fewer, binding to pocket A in WT FabB is significantly more stable ($p < 0.05$). For chains longer than 10 carbons pocket B is either statistically equivalent or significantly more stable than pocket A. We cannot perform a quantitative comparison to experimental data as there is no available experimental data on binding free energy for the covalent intermediate or on the relative occupancy of pocket A or pocket B. Importantly, these calculations also do not consider the relative stability of the ACP-complex configuration which would be necessary for direct quantitative comparison to experimental biding free energies. Given the complexity of interpreting the accuracy of our free energy estimates from the available experimental data for WT FabB alone, we compare the free energy surface between the WT and G107M variant of FabB.

\begin{figure}[H]
    \centering
    \includegraphics[width=0.8\linewidth]{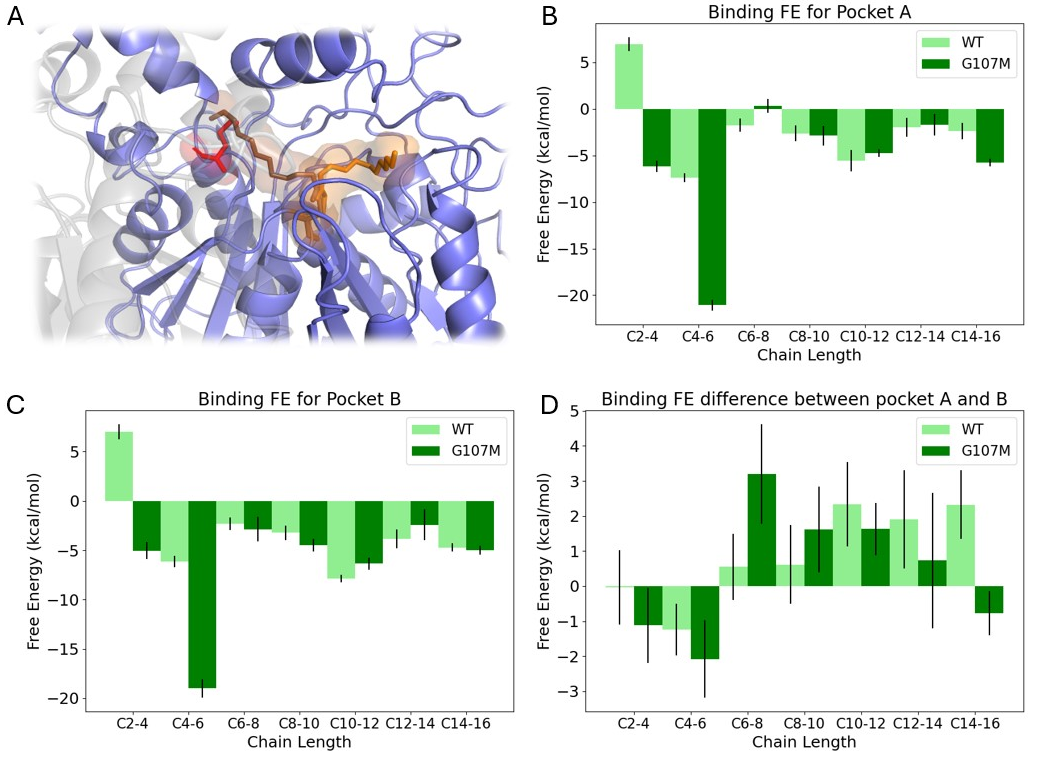}
    \begin{singlespace}
    \caption{(A) The G107M mutation (red) blocks access to a region of pocket A which destabilizes chain binding to this pocket, but has little direct affect on pocket B. We show centroids for the C10 to C12 transformation for the chain in pocket A (brown) and pocket B (orange). To assess the effect of this mutation on the binding complex, we compare the binding free energy differences by chain length for the WT and G107M variant in pocket A (B) and pocket B (C). (D) To highlight the difference in stability between the two pockets we then compare the difference in individual relative binding free energies between pocket A and B for all chain lengths for the WT and G107M variant. There is a clear bias in the mutant FabB variant against the growth of the acyl chain from 6 to 8 and from 8 to 10 carbons.}
    \label{fig:FabB_G107M}
    \end{singlespace}
\end{figure}

In \textit{E. coli}, FabB and FabF are homologs that show different responses to mutations. In FabF, I108M prevents the synthesis of acyl chains longer than 8 carbons; in FabB, G107M (the analogous mutation) is more permissive but still shifts product profiles toward shorter chains~\cite{mainsKetosynthaseMutantsEnable2023a}. These effects suggest that only FabB can access pocket B, which permits binding of longer chains when pocket A is obstructed. (Figure \ref{fig:FabB_G107M}A). Focusing on FabB, we applied the MT-REXEE approach to measure binding free energy estimates for all chain lengths in pockets A and B to determine how G107M affects the stability of the covalent intermediate for different chain lengths (Figure \ref{fig:FabB_G107M}B--C). It should be noted that we do not present a comparison of the overall binding free energy for each transformation. This quantity can easily be calculated from the simulations, but our goal is to quantify the relative stability of the two possible complexes with and without the mutation.  In addition, the experimental data which we have for comparison in this work is derived from enzyme activity assays which would not provide us with a direct quantity to compare to an overall binding affinity. 

Interestingly, the G107M mutation appears to stabilize the binding of C2--6 chain lengths, as shown by the negative free energy difference between the mutations, shown in Figure \ref{fig:FabB_G107M}D. The larger residue at position 107 forms additional non-bonded interactions within the pocket as well as with the short acyl chains, which may explain this stabilizing affect. For the C6 to C8 transformation, however, we see a significant destabilization of pocket A in the G107M variant, relative to WT, as shown by the positive free energy difference \ref{fig:FabB_G107M}D. This difference is particularly evident for pocket A but not for pocket B, as shown by the difference in relative binding free energy between pockets A and B ($\Delta G_A - \Delta G_B$). In the WT variant, chain growth into pocket A is more stable than pocket B until 6 carbons, the two pockets are statistically identical until 10 carbons, and then for 12 carbons or greater, pocket B is more stable for each chain growth step. However, for G107M, pocket B is significantly more stable while growing the chain from 6 to 10 carbons (Figure \ref{fig:FabB_G107M}D). This stark difference in relative binding free energies between pockets A and B coincides with the chain length at which binding in pocket A is predicted to be destabilized based on experimental evidence. 

\section{Discussion and Conclusions}
The MT-REXEE method increases conformational sampling for sets of alchemical transformations in which prohibitively high free energy barriers are present at only a subset of the states of interest. In this paper, we used the FabB-acyl complex to demonstrate the potential of MT-REXEE to increase conformational sampling. In the FabB-acyl complex, the acyl chain could access two distinct pockets, consistent with previous crystallographic analyses~\cite{jiangHiddenBindingPocket2026}. Pocket B was inaccessible using standard MD simulations initiated in pocket A for any chain length longer than 4 carbons. MT-REXEE allowed for significantly enhanced sampling between the two pockets within a single trajectory for each transformation. This enhanced sampling enabled quantification of the difference in binding free energy surfaces between pocket A and pocket B for the G107M FabB variant, which shifts product profiles toward short chains but still permits medium chain binding.

This work also highlights the new exchange method, random range (RR), in which conformational exchanges are allowed at any frame within a given percentage of the trajectory rather than the last frame of an iteration. This improvement removes the necessity of including redundant or up-weighted end states which was previously necessary. We verified that this novel exchange method does not introduce additional error into the calculations through comparison of binding free energy estimates to MUP1 using both EE and MT-REXEE using the previous exhaustive method. All three methods produce statistically equivalent binding FE estimates, but we observe significantly faster convergence using MT-REXEE with the RR method. The RR procedure significantly reduces the round-trip times in alchemical space while also increasing relative sampling of intermediate $\lambda$ states with the same trajectory length. Decreasing the round-trip time increases the efficiency of the MT-REXEE method, which is of growing importance when simulating larger systems like that of the FabB--acyl complex. There is a trade-off when using the RR method in which an average of $\sim$25\% of each iteration's sampling (for the settings used in this study) is not included in the trajectories leading to conformational swaps. This sampling is still included in FE and trajectory analysis and is thus not lost, but will reduce the degree of decorrelation of the pre-swap configurations because of decreased total sampling time along continuous trajectories.

For the FabB-acyl complex, we simulated the covalent intermediate state. This state was selected to simplify the test system, compared to the FabB--acyl-ACP complex, and also to demonstrate the method's potential application to covalent ligand systems. Through this example system, we demonstrate the ability for MT-REXEE simulations to sample alternate conformational states without known prior definitions. Crystallography data did support the presence of an alternate binding pocket in the FabB--acyl-ACP complex, but it was not known what form, if any, this alternate pocket would arise in the covalent intermediate. MT-REXEE simulations sampled two previously unknown pockets in the covalent intermediate including one which resembled that of pocket B in the ACP complex and one novel conformation (pocket C). This unbiased alternate pocket discovery potential is the primary novel advantage to the MT-REXEE method.

The large highly flexible FabB-acyl system does pose important unsolved challenges for the MT-REXEE method. The Wang-Landau weight equilibration times can exceed 100 ns thus requiring significant additional computational resources due to the complex nature of the FE landscape. This problem could be mitigated by implementing educated initial guesses for weights and using adaptive sampling methods like SAMS~\cite{tanOptimallyAdjustedMixture2017}. Additionally, handling multiple simulations that need to be exchanged on the order of minutes, although removing complexity from simulation code, still requires a complex workflow that is not optimally managed by sub-process calls in Python.  Workflow software improvements for robustly managing large numbers of biomolecular simulations at scale would allow for simulations to be run much more easily~\cite{Balasubramanian:SCS:2020}. The current framework is functionally limited to running on a single CPU node due to significant delays caused by frequent inter-node communication and is thus hardware-limited. 

Despite these limitations, MT-REXEE provides a unique methodology for completely unbiased sampling of alternate conformational states. The intended application for large-flexible systems is not high throughput analysis but rather in-depth analysis of challenging systems for which other methods are either insufficient or require assumptions for which sufficient data is not available to support. This includes systems for which alternate binding modes are suspected but not known. This was the case for the FabB-acyl complex system, where MT-REXEE simulations were able to provide valuable insights on occupancy of alternate pockets and explain unexpected mutational effects.

\section*{Author Contributions}
A.J.F, M.R.S., and J.M.F conceptualized the project and designed the methodology. A J. F. implemented the modifications to sampling procedures in Python. Experiments were performed and analyzed by A.J.F.. A.J.F. wrote the original manuscript draft; M.R.S. and J.M.F edited and reviewed the manuscript.  M.R.S. supervised the project and obtained resources.

\section*{Supporting Information}
The mathematical details of the non-additive nature of the Gibbs free energy differences and diagrams detailing $\lambda$ window sampling and FE differences within cycles are included within the SI. The SI also includes details regarding binding pocket identification, individual transformation equilibration times, individual transformation FE differences, details regarding simulation replica swapping, reference states structures for acyl-ACP, and $\lambda$ state overlap matrices.

\begin{acknowledgement}
This study is based on work supported by grant OAC-1835720 from the National Science Foundation, grants R01GM123296 and R35GM158359 from the National Institutes of General Medical Sciences (A.J.F. and M.R.S.), and grant DE-SC0023142 from the Department of Energy (J.M.F.). This work utilized the Alpine high performance computing resource at the University of Colorado Boulder. Alpine is jointly funded by the University of Colorado Boulder, the University of Colorado Anschutz, and Colorado State University. The content is solely the responsibility of the authors and does not necessarily represent the official views of the National Institutes of Health, National Science Foundation, or Department of Energy.
\end{acknowledgement}

\clearpage
\bibliography{citations}

@article{decescoCovalentInhibitorsDesign2017,
  title = {Covalent Inhibitors Design and Discovery},
  author = {De Cesco, Stephane and Kurian, Jerry and Dufresne, Caroline and Mittermaier, Anthony K. and Moitessier, Nicolas},
  year = {2017},
  month = sep,
  journal = {European Journal of Medicinal Chemistry},
  volume = {138},
  pages = {96--114},
  issn = {1768-3254},
  langid = {english},
  pmid = {28651155}
}

@article{marcellaStructureHighAffinity2017,
  title = {Structure, High Affinity, and Negative Cooperativity of the Escherichia coli Holo-(Acyl Carrier Protein):Holo-(Acyl Carrier Protein) Synthase Complex},
  author = {Marcella, Aaron M. and Culbertson, Sannie J. and Shogren-Knaak, Michael A. and Barb, Adam W},
  year = {2017},
  month = nov,
  journal = {Journal of Molecular Biology},
  volume = {429},
  number={23},
}

@article{priceStructureOfBeta2001,
  title = {Structure of beta-ketoacyl-[acyl carrier protein] reductase from Escherichia coli: negative cooperativity and its structural basis},
  author = {Price, A. C. and Zhang, Y. M. and White, S. W.},
  year = {2001},
  month = oct,
  journal = {Biochemistry},
  volume = {40},
  number={43},
}

@article{boreschRoleBondedEnergy2002,
  title = {The {{Role}} of {{Bonded Energy Terms}} in {{Free Energy Simulations}} - {{Insights}} from {{Analytical Results}}},
  author = {Boresch, Stefan},
  year = {2002},
  month = jan,
  journal = {Molecular Simulation},
  volume = {28},
  number = {1-2},
  pages = {13--37},
  publisher = {Taylor \& Francis},
  issn = {0892-7022},
  doi = {10.1080/08927020211969}
}

@article{zwierWESTPAInteroperableHighly2015,
  title = {{{WESTPA}}: {{An Interoperable}}, {{Highly Scalable Software Package}} for {{Weighted Ensemble Simulation}} and {{Analysis}}},
  shorttitle = {{{WESTPA}}},
  author = {Zwier, Matthew C. and Adelman, Joshua L. and Kaus, Joseph W. and Pratt, Adam J. and Wong, Kim F. and Rego, Nicholas B. and Su{\'a}rez, Ernesto and Lettieri, Steven and Wang, David W. and Grabe, Michael and Zuckerman, Daniel M. and Chong, Lillian T.},
  year = {2015},
  month = feb,
  journal = {Journal of Chemical Theory and Computation},
  volume = {11},
  number = {2},
  pages = {800--809},
  publisher = {American Chemical Society},
  issn = {1549-9618},
  doi = {10.1021/ct5010615}
}

@article{schaeferRecentAdvancesCovalent2023,
  title = {Recent {{Advances}} in {{Covalent Drug Discovery}}},
  author = {Schaefer, Daniel and Cheng, Xinlai},
  year = {2023},
  month = may,
  journal = {Pharmaceuticals},
  volume = {16},
  number = {5},
  pages = {663},
  publisher = {Multidisciplinary Digital Publishing Institute},
  issn = {1424-8247},
  doi = {10.3390/ph16050663},
  copyright = {http://creativecommons.org/licenses/by/3.0/},
  langid = {english}
}

@article{singhResurgenceCovalentDrugs2011,
  title = {The Resurgence of Covalent Drugs},
  author = {Singh, Juswinder and Petter, Russell C. and Baillie, Thomas A. and Whitty, Adrian},
  year = {2011},
  month = apr,
  journal = {Nature Reviews Drug Discovery},
  volume = {10},
  number = {4},
  pages = {307--317},
  publisher = {Nature Publishing Group},
  issn = {1474-1784},
  doi = {10.1038/nrd3410},
  copyright = {2011 Springer Nature Limited},
  langid = {english}
}

@article{wuAlchemlybSimpleAlchemistry2024,
  title = {Alchemlyb: The Simple Alchemistry Library},
  shorttitle = {Alchemlyb},
  author = {Wu, Zhiyi and Dotson, David L. and Alibay, Irfan and Allen, Bryce K. and Barhaghi, Mohammad Soroush and H{\'e}nin, J{\'e}r{\^o}me and Joseph, Thomas T. and Kenney, Ian M. and Lee, Hyungro and Li, Haoxi and Lim, Victoria and Liu, Shuai and Marson, Domenico and Merz, Pascal T. and Schlaich, Alexander and Mobley, David and Shirts, Michael R. and Beckstein, Oliver},
  year = {2024},
  month = sep,
  journal = {Journal of Open Source Software},
  volume = {9},
  number = {101},
  pages = {6934},
  issn = {2475-9066},
  doi = {10.21105/joss.06934},
  langid = {english}
}

@article{choderaAlchemicalFreeEnergy2011a,
  title = {Alchemical Free Energy Methods for Drug Discovery: {{Progress}} and Challenges},
  shorttitle = {Alchemical Free Energy Methods for Drug Discovery},
  author = {Chodera, John D. and Mobley, David L. and Shirts, Michael R. and Dixon, Richard W. and Branson, Kim and Pande, Vijay S.},
  year = {2011},
  month = apr,
  journal = {Curr Opin Struct Biol},
  volume = {21},
  number = {2},
  pages = {150--160},
  issn = {0959-440X},
  doi = {10.1016/j.sbi.2011.01.011},
  urldate = {2024-06-27},
  pmcid = {PMC3085996},
  pmid = {21349700}
}

@article{fleckDummyAtomsAlchemical2021a,
  title = {Dummy {{Atoms}} in {{Alchemical Free Energy Calculations}}},
  author = {Fleck, Markus and Wieder, Marcus and Boresch, Stefan},
  year = {2021},
  month = jul,
  journal = {J. Chem. Theory Comput.},
  volume = {17},
  number = {7},
  pages = {4403--4419},
  publisher = {American Chemical Society},
  issn = {1549-9618},
  doi = {10.1021/acs.jctc.0c01328},
  urldate = {2024-08-16}
}

@article{mueggeRecentAdvancesAlchemical2023,
  title = {Recent {{Advances}} in {{Alchemical Binding Free Energy Calculations}} for {{Drug Discovery}}},
  author = {Muegge, Ingo and Hu, Yuan},
  year = {2023},
  month = mar,
  journal = {ACS Med. Chem. Lett.},
  volume = {14},
  number = {3},
  pages = {244--250},
  publisher = {American Chemical Society},
  doi = {10.1021/acsmedchemlett.2c00541},
  urldate = {2024-06-27}
}

@article{yorkModernAlchemicalFree2023,
  title = {Modern {{Alchemical Free Energy Methods}} for {{Drug Discovery Explained}}},
  author = {York, Darrin M.},
  year = {2023},
  month = nov,
  journal = {ACS Phys. Chem Au},
  volume = {3},
  number = {6},
  pages = {478--491},
  publisher = {American Chemical Society},
  doi = {10.1021/acsphyschemau.3c00033},
  urldate = {2024-06-27}
}

@article{meyBestPracticesAlchemical2020a,
  title = {Best {{Practices}} for {{Alchemical Free Energy Calculations}} [{{Article}} v1.0]},
  author = {Mey, Antonia S. J. S. and Allen, Bryce K. and McDonald, Hannah E. Bruce and Chodera, John D. and Hahn, David F. and Kuhn, Maxmillian and Michel, Julien and Mobley, David L. and Naden, Levi N. and Prasad, Samarjeet and Rizzi, Andrea and Scheen, Jenke and Shirts, Michael R. and Tresadern, Gary and Xu, Huafeng},
  year = {2020},
  month = dec,
  journal = {Living Journal of Computational Molecular Science},
  volume = {2},
  number = {1},
  pages = {18378--18378},
  issn = {2575-6524},
  doi = {10.33011/livecoms.2.1.18378},
  urldate = {2024-11-11},
  copyright = {Copyright (c) 2020},
  langid = {english}
}

@article{papadourakisAlchemicalFreeEnergy2023,
  title = {Alchemical {{Free Energy Calculations}} on {{Membrane-Associated Proteins}}},
  author = {Papadourakis, Michail and Sinenka, Hryhory and Matricon, Pierre and H{\'e}nin, J{\'e}r{\^o}me and Brannigan, Grace and {P{\'e}rez-Benito}, Laura and Pande, Vineet and {van Vlijmen}, Herman and {de Graaf}, Chris and Deflorian, Francesca and Tresadern, Gary and Cecchini, Marco and Cournia, Zoe},
  year = {2023},
  month = nov,
  journal = {Journal of Chemical Theory and Computation},
  volume = {19},
  number = {21},
  pages = {7437--7458},
  publisher = {American Chemical Society},
  issn = {1549-9618},
  doi = {10.1021/acs.jctc.3c00365}
}

@article{abelCriticalReviewValidation2017,
  title = {A {{Critical Review}} of {{Validation}}, {{Blind Testing}}, and {{Real- World Use}} of {{Alchemical Protein-Ligand Binding Free Energy Calculations}}},
  author = {Abel, Robert and Wang, Lingle and Mobley, David L. and Friesner, Richard A.},
  year = {2017},
  month = sep,
  journal = {Current Topics in Medicinal Chemistry},
  volume = {17},
  number = {23},
  pages = {2577--2585},
  doi = {10.2174/1568026617666170414142131}
}

@article{jiangReducedFreeEnergy2018,
  title = {Reduced {{Free Energy Perturbation}}/{{Hamiltonian Replica Exchange Molecular Dynamics Method}} with {{Unbiased Alchemical Thermodynamic Axis}}},
  author = {Jiang, Wei and Thirman, Jonathan and Jo, Sunhwan and Roux, Beno{\^i}t},
  year = {2018},
  month = oct,
  journal = {J. Phys. Chem. B},
  volume = {122},
  number = {41},
  pages = {9435--9442},
  publisher = {American Chemical Society},
  issn = {1520-6106},
  doi = {10.1021/acs.jpcb.8b03277},
  urldate = {2024-07-02}
}

@article{laserraAlchemicalFreeEnergy2022,
  title = {Alchemical {{Free Energy Calculations}} to {{Investigate Protein}}--{{Protein Interactions}}: The {{Case}} of the {{CDC42}}/{{PAK1 Complex}}},
  shorttitle = {Alchemical {{Free Energy Calculations}} to {{Investigate Protein}}--{{Protein Interactions}}},
  author = {La Serra, Maria Antonietta and Vidossich, Pietro and Acquistapace, Isabella and Ganesan, Anand K. and De Vivo, Marco},
  year = {2022},
  month = jun,
  journal = {J. Chem. Inf. Model.},
  volume = {62},
  number = {12},
  pages = {3023--3033},
  publisher = {American Chemical Society},
  issn = {1549-9596},
  doi = {10.1021/acs.jcim.2c00348},
  urldate = {2024-09-09}
}

@article{zhangExpandedEnsembleMethods2021,
  title = {Expanded {{Ensemble Methods Can}} Be {{Used}} to {{Accurately Predict Protein-Ligand Relative Binding Free Energies}}},
  author = {Zhang, Si and Hahn, David F. and Shirts, Michael R. and Voelz, Vincent A.},
  year = {2021},
  month = oct,
  journal = {J. Chem. Theory Comput.},
  volume = {17},
  number = {10},
  pages = {6536--6547},
  publisher = {American Chemical Society},
  issn = {1549-9618},
  doi = {10.1021/acs.jctc.1c00513},
  urldate = {2024-06-30}
}

@article{pflegerRecentProgressSynthesis2023,
  title = {Recent Progress in the Synthesis of Advanced Biofuel and Bioproducts},
  author = {Pfleger, Brian F and Takors, Ralf},
  year = {2023},
  month = apr,
  journal = {Current Opinion in Biotechnology},
  volume = {80},
  pages = {102913},
  issn = {0958-1669},
  doi = {10.1016/j.copbio.2023.102913}
}

@article{singhProductionBiofuelsOptions2022,
  title = {Production of Biofuels Options by Contribution of Effective and Suitable Enzymes: {{Technological}} Developments and Challenges},
  shorttitle = {Production of Biofuels Options by Contribution of Effective and Suitable Enzymes},
  author = {Singh, Renu and Langyan, Sapna and Rohtagi, Bharti and Darjee, Sibananda and Khandelwal, Ashish and Shrivastava, Manoj and Kothari, Richa and Mohan, Har and Raina, Shubham and Kaur, Japleen and Singh, Anita},
  year = {2022},
  month = jan,
  journal = {Materials Science for Energy Technologies},
  volume = {5},
  pages = {294--310},
  issn = {2589-2991},
  doi = {10.1016/j.mset.2022.05.001}
}

@article{buyachuihanHowAcylCarrier2024,
  title = {How {{Acyl Carrier Proteins}} ({{ACPs}}) {{Direct Fatty Acid}} and {{Polyketide Biosynthesis}}},
  author = {Buyachuihan, Lynn and Stegemann, Franziska and Grininger, Martin},
  year = {2024},
  month = jan,
  journal = {Angewandte Chemie (International Ed. in English)},
  volume = {63},
  number = {4},
  pages = {e202312476},
  issn = {1521-3773},
  doi = {10.1002/anie.202312476},
  langid = {english},
  pmid = {37856285}
}

@article{chanCurrentUnderstandingFatty2010,
  title = {Current Understanding of Fatty Acid Biosynthesis and the Acyl Carrier Protein},
  author = {Chan, David I. and Vogel, Hans J.},
  year = {2010},
  month = aug,
  journal = {The Biochemical Journal},
  volume = {430},
  number = {1},
  pages = {1--19},
  issn = {1470-8728},
  doi = {10.1042/BJ20100462},
  langid = {english},
  pmid = {20662770}
}

@article{mainsKetosynthaseMutantsEnable2023a,
  title = {Ketosynthase Mutants Enable Short-Chain Fatty Acid Biosynthesis in {{{\emph{E}}}}{\emph{. Coli}}},
  author = {Mains, Kathryn and Fox, Jerome M.},
  year = {2023},
  month = may,
  journal = {Metabolic Engineering},
  volume = {77},
  pages = {118--127},
  issn = {1096-7176},
  doi = {10.1016/j.ymben.2023.03.008}
}

@article{abergDeterminationSolvationFree2004,
  title = {Determination of Solvation Free Energies by Adaptive Expanded Ensemble Molecular Dynamics},
  author = {{\AA}berg, K. Magnus and Lyubartsev, Alexander P. and Jacobsson, Sven P. and Laaksonen, Aatto},
  year = {2004},
  month = feb,
  journal = {The Journal of Chemical Physics},
  volume = {120},
  number = {8},
  pages = {3770--3776},
  issn = {0021-9606},
  doi = {10.1063/1.1642601},
  urldate = {2024-11-11}
}

@article{you_evaluating_2017,
	title = {Evaluating the accuracy of the umbrella sampling plots with different dissociation paths, conformational changes, and structure preparation},
	rights = {© 2017, Posted by Cold Spring Harbor Laboratory. The copyright holder for this pre-print is the author. All rights reserved. The material may not be redistributed, re-used or adapted without the author's permission.},
	url = {https://www.biorxiv.org/content/10.1101/169532v1},
	doi = {10.1101/169532},
	pages = {169532},
	journal = {{bioRxiv}},
	author = {You, Wanli and Tang, Zhiye and Chang, Chia-en},
	urldate = {2019-01-31},
	date = {2017-07-28},
        month = {January},
        year = {2017},
	langid = {english},
}

@article{min_convergence_2007,
	title = {On the convergence improvement in the metadynamics simulations: A Wang-Landau recursion approach},
	volume = {126},
	issn = {00219606},
	url = {http://link.aip.org/link/JCPSA6/v126/i19/p194104/s1&Agg=doi},
	doi = {10.1063/1.2731769},
	pages = {194104},
	number = {19},
	journal = {The Journal of Chemical Physics},
	shortjournal = {J. Chem. Phys.},
	author = {Min, Donghong and Liu, Yusong and Carbone, Irina and Yang, Wei},
	year = {2007},
        month = {May},
}

@article{torrie_nonphysical_1977,
	title = {Nonphysical sampling distributions in Monte Carlo free-energy estimation: Umbrella sampling},
	volume = {23},
	issn = {0021-9991},
	url = {https://www.sciencedirect.com/science/article/pii/0021999177901218},
	doi = {10.1016/0021-9991(77)90121-8},
	shorttitle = {Nonphysical sampling distributions in Monte Carlo free-energy estimation},
	pages = {187--199},
	number = {2},
	journal = {Journal of Computational Physics},
	shortjournal = {Journal of Computational Physics},
	author = {Torrie, G. M. and Valleau, J. P.},
	urldate = {2024-09-16},
	date = {1977-02-01},
        year = {1977},
        month = {February},
}

@article{hsu_alchemical_2023,
	title = {Alchemical Metadynamics: Adding Alchemical Variables to Metadynamics to Enhance Sampling in Free Energy Calculations},
	volume = {19},
	issn = {1549-9618},
	url = {https://doi.org/10.1021/acs.jctc.2c01258},
	doi = {10.1021/acs.jctc.2c01258},
	shorttitle = {Alchemical Metadynamics},
	pages = {1805--1817},
	number = {6},
	journal = {Journal of Chemical Theory and Computation},
	shortjournal = {J. Chem. Theory Comput.},
	author = {Hsu, Wei-Tse and Piomponi, Valerio and Merz, Pascal T. and Bussi, Giovanni and Shirts, Michael R.},
	urldate = {2024-06-30},
	date = {2023-03-28},
        month = {March},
        year = {2023},
}

@article{robo_fast_2023,
	title = {Fast free energy estimates from $\lambda$-dynamics with bias-updated Gibbs sampling},
	volume = {14},
	rights = {2023 The Author(s)},
	issn = {2041-1723},
	url = {https://www.nature.com/articles/s41467-023-44208-9},
	doi = {10.1038/s41467-023-44208-9},
	pages = {8515},
	number = {1},
	journal = {Nature Communications},
	shortjournal = {Nat Commun},
	author = {Robo, Michael T. and Hayes, Ryan L. and Ding, Xinqiang and Pulawski, Brian and Vilseck, Jonah Z.},
	urldate = {2024-07-02},
	date = {2023-12-21},
        month = {December},
        year = {2023},
	langid = {english},
}

@article{friedmanMultipleTopologyReplica2025,
  title = {Multiple {{Topology Replica Exchange}} of {{Expanded Ensembles}} for {{Multidimensional Alchemical Calculations}}},
  author = {Friedman, Anika J. and Hsu, Wei-Tse and Shirts, Michael R.},
  year = {2025},
  month = jan,
  journal = {Journal of Chemical Theory and Computation},
  publisher = {American Chemical Society},
  issn = {1549-9618},
  doi = {10.1021/acs.jctc.4c01268},
}

@misc{bauerGROMACS20225Source2023,
  title = {{{GROMACS}} 2022.5 {{Source}} Code},
  author = {Bauer, Paul and Hess, Berk and Lindahl, Erik},
  year = {2023},
  month = feb,
  doi = {10.5281/zenodo.7586780},
  urldate = {2024-09-09},
  howpublished = {Zenodo}
}

@article{bernettiPressureControlUsing2020a,
  title = {Pressure Control Using Stochastic Cell Rescaling},
  author = {Bernetti, Mattia and Bussi, Giovanni},
  year = {2020},
  month = sep,
  journal = {The Journal of Chemical Physics},
  volume = {153},
  number = {11},
  pages = {114107},
  issn = {0021-9606},
  doi = {10.1063/5.0020514},
  urldate = {2024-08-16}
}

@article{bussiCanonicalSamplingVelocity2007a,
  title = {Canonical Sampling through Velocity Rescaling},
  author = {Bussi, Giovanni and Donadio, Davide and Parrinello, Michele},
  year = {2007},
  month = jan,
  journal = {J Chem Phys},
  volume = {126},
  number = {1},
  pages = {014101},
  issn = {0021-9606},
  doi = {10.1063/1.2408420},
  langid = {english},
  pmid = {17212484}
}

@article{gapsysPmxWebserverUser2017a,
  title = {Pmx {{Webserver}}: {{A User Friendly Interface}} for {{Alchemistry}}},
  shorttitle = {Pmx {{Webserver}}},
  author = {Gapsys, Vytautas and {de Groot}, Bert L.},
  year = {2017},
  month = feb,
  journal = {J. Chem. Inf. Model.},
  volume = {57},
  number = {2},
  pages = {109--114},
  publisher = {American Chemical Society},
  issn = {1549-9596},
  doi = {10.1021/acs.jcim.6b00498},
  urldate = {2024-08-16}
}

@article{heFastHighqualityCharge2020,
  title = {A Fast and High-Quality Charge Model for the next Generation General {{AMBER}} Force Field},
  author = {He, Xibing and Man, Viet H. and Yang, Wei and Lee, Tai-Sung and Wang, Junmei},
  year = {2020},
  month = sep,
  journal = {The Journal of Chemical Physics},
  volume = {153},
  number = {11},
  pages = {114502},
  issn = {0021-9606},
  doi = {10.1063/5.0019056},
  urldate = {2024-08-16}
}

@article{jorgensenComparisonSimplePotential1983,
  title = {Comparison of Simple Potential Functions for Simulating Liquid Water},
  author = {Jorgensen, William L. and Chandrasekhar, Jayaraman and Madura, Jeffry D. and Impey, Roger W. and Klein, Michael L.},
  year = {1983},
  month = jul,
  journal = {The Journal of Chemical Physics},
  volume = {79},
  number = {2},
  pages = {926--935},
  issn = {0021-9606},
  doi = {10.1063/1.445869},
  urldate = {2024-09-18}
}

@article{maierFf14SBImprovingAccuracy2015a,
  title = {{{ff14SB}}: {{Improving}} the {{Accuracy}} of {{Protein Side Chain}} and {{Backbone Parameters}} from {{ff99SB}}},
  shorttitle = {{{ff14SB}}},
  author = {Maier, James A. and Martinez, Carmenza and Kasavajhala, Koushik and Wickstrom, Lauren and Hauser, Kevin E. and Simmerling, Carlos},
  year = {2015},
  month = aug,
  journal = {J. Chem. Theory Comput.},
  volume = {11},
  number = {8},
  pages = {3696--3713},
  publisher = {American Chemical Society},
  issn = {1549-9618},
  doi = {10.1021/acs.jctc.5b00255},
  urldate = {2024-08-16}
}

@article{tanOptimallyAdjustedMixture2017,
  title = {Optimally {{Adjusted Mixture Sampling}} and {{Locally Weighted Histogram Analysis}}},
  author = {Tan, Zhiqiang},
  year = 2017,
  month = jan,
  journal = {Journal of Computational and Graphical Statistics},
  volume = {26},
  number = {1},
  pages = {54--65},
  issn = {1061-8600, 1537-2715},
  doi = {10.1080/10618600.2015.1113975},
  urldate = {2026-02-15},
  langid = {english}
}

@article{Balasubramanian:SCS:2020,
  title = {Adaptive {{Ensemble Biomolecular Applications}} at {{Scale}}},
  author = {Balasubramanian, Vivek and Jensen, Travis and Turilli, Matteo and Kasson, Peter and Shirts, Michael and Jha, Shantenu},
  year = 2020,
  month = mar,
  journal = {SN Computer Science},
  volume = {1},
  number = {2},
  pages = {104},
}

@article{jiangHiddenBindingPocket2026,
  title = {A {{Hidden Binding Pocket}} in the {$\beta$}- Ketoacyl-{{ACP Synthase FabB}}},
  author = {Jiang, Ziran and Friedman, Anika J. and Thompson, Annette and Andrzejewski, Samuel J. and Mains, Kathryn and Sankaran, Banumathi and Burkart, Michael Jane and Shirts, Michael R. and Fox, Jerome Michael},
  year = {2026},
  month = feb,
  primaryclass = {New Results},
  pages = {2026.02.26.708327},
  publisher = {bioRxiv},
  journal = {bioRxiv},
  issn = {2692-8205},
  doi = {10.64898/2026.02.26.708327},
  urldate = {2026-03-01},
  archiveprefix = {bioRxiv},
  chapter = {New Results},
  copyright = {{\copyright} 2026, Posted by openRxiv. This pre-print is available under a Creative Commons License (Attribution-NonCommercial-NoDerivs 4.0 International), CC BY-NC-ND 4.0, as described at http://creativecommons.org/licenses/by-nc-nd/4.0/},
  langid = {english},
}

@article{mindreboGatingMechanismElongating2020a,
  title = {Gating Mechanism of Elongating {$\beta$}-Ketoacyl-{{ACP}} Synthases},
  author = {Mindrebo, Jeffrey T. and Patel, Ashay and Kim, Woojoo E. and Davis, Tony D. and Chen, Aochiu and Bartholow, Thomas G. and La Clair, James J. and McCammon, J. Andrew and Noel, Joseph P. and Burkart, Michael D.},
  year = {2020},
  month = apr,
  journal = {Nature Communications},
  volume = {11},
  number = {1},
  pages = {1727},
  publisher = {Nature Publishing Group},
  issn = {2041-1723},
  doi = {10.1038/s41467-020-15455-x},
  urldate = {2026-03-01},
  copyright = {2020 The Author(s)},
  langid = {english}
}

@article{daltonMedicinalChemistryPerspective2025,
  title = {A {{Medicinal Chemistry Perspective}} on {{FDA-Approved Small Molecule Drugs}} with a {{Covalent Mechanism}} of {{Action}}},
  author = {Dalton, Samuel E. and Di Pietro, Ornella and Hennessy, Elisabeth},
  year = {2025},
  month = feb,
  journal = {Journal of Medicinal Chemistry},
  volume = {68},
  number = {3},
  pages = {2307--2313},
  publisher = {American Chemical Society},
  issn = {0022-2623},
  doi = {10.1021/acs.jmedchem.4c02661},
  urldate = {2026-03-01}
}

@article{ruppeKineticRationaleFunctional2020,
  title = {A Kinetic Rationale for Functional Redundancy in Fatty Acid Biosynthesis},
  author = {Ruppe, Sophia and Mains, Kathryn and Fox, Jerome M.},
  year = {2020},
  month = sep,
  journal = {Proceedings of the National Academy of Sciences},
  volume = {117},
  number = {38},
  pages = {23557--23564},
  publisher = {Proceedings of the National Academy of Sciences},
  doi = {10.1073/pnas.2013924117},
  urldate = {2026-03-01},
  langid = {english}
}

@article{peoplesKineticFrameworkModeling2022,
  title = {A Kinetic Framework for Modeling Oleochemical Biosynthesis in {{Escherichia}} Coli},
  author = {Peoples, Jackson and Ruppe, Sophia and Mains, Kathryn and Cipriano, Elia C. and Fox, Jerome M.},
  year = 2022,
  journal = {Biotechnology and Bioengineering},
  volume = {119},
  number = {11},
  pages = {3149--3161},
  issn = {1097-0290},
  doi = {10.1002/bit.28209},
  urldate = {2026-06-28},
  copyright = {\copyright{} 2022 Wiley Periodicals LLC.},
  langid = {english}
}

\newpage
\renewcommand{\thetable}{S\arabic{table}}
\renewcommand{\thefigure}{S\arabic{figure}}
\renewcommand{\thesection}{S\arabic{section}}
\renewcommand{\thepage}{S\arabic{page}}
\setcounter{figure}{0}

\section*{Supplemental Information}
\subsection*{Non-additive nature of $\Delta A_{ij, A}$ and $\Delta A_{ij, B}$}\label{SI:fe_add}
The Gibbs FE differences are not directly additive $\Delta G_{ij} \neq \Delta G_{ij, A} + \Delta G_{ij, B}$ as this would assume equal contributions from both configurational states. To get a direct expression for the free energy difference between states $i$ and $j$ regardless of configurational state ($\Delta G_{ij}$) in terms of $\Delta G_{ij, A}$ and $\Delta G_{ij, B}$ we start by rearranging equation \ref{eq:new_core} to give us equation \ref{eq:GA} and \ref{eq:GB}. For simplicity's sake, we utilize the variables $A_i$, $A_j$, $B_i$, and $B_j$ as defined in equations \ref{eq:GA} and \ref{eq:GB}.
\begin{equation}\label{eq:GA}
    \frac{A_j}{A_i} = \frac{\int_{\Gamma_{j,A}}e^{-\frac{U_j(\vec{q})}{k_BT}}d\vec{q}}{\int_{\Gamma_{i,A}}e^{-\frac{U_i(\vec{q})}{k_BT}}d\vec{q}} = e^{\frac{-\Delta A_{ij, A}}{k_BT}}
\end{equation}
\begin{equation}\label{eq:GB}
    \frac{B_j}{B_i} = \frac{\int_{\Gamma_{j,B}}e^{-\frac{U_j(\vec{q})}{k_BT}}d\vec{q}}{\int_{\Gamma_{i,B}}e^{-\frac{U_i(\vec{q})}{k_BT}}d\vec{q}} = e^{\frac{-\Delta A_{ij, B}}{k_BT}}
\end{equation}
With this reduced representation we can rewrite equation \ref{eq:core_FE} as $\Delta G_{ij} = -k_BT\ln{\frac{A_i + B_i}{A_j + B_j}}$. When we substitute in the values for Equations \ref{eq:GA} and \ref{eq:GB} we get 
\begin{equation}
    \Delta G_{ij} = -k_BT\ln{\frac{e^{\frac{-\Delta A_{ij, A}}{k_BT}} + e^{\frac{-\Delta A_{ij, B}}{k_BT}}}{1 + \frac{A_i}{B_i}}}
\end{equation}
which can finally be reduced to Equation \ref{eq:dG_sum} in which the final term represents the adjustment that must be made to account for uneven relative state weighting.
\begin{equation}\label{eq:dG_sum}
    \Delta G_{ij} = \Delta G_{ij, A} + \Delta G_{ij, B} + k_BT\ln{(1+\frac{\int_{\Gamma_{i,A}}e^{-\frac{U_i(\vec{q})}{k_BT}}d\vec{q}}{\int_{\Gamma_{i,B}}e^{-\frac{U_i(\vec{q})}{k_BT}}d\vec{q}})}
\end{equation}
We can directly compute $\Delta G_{ij}$ from Equation \ref{eq:core_FE} with all trajectory frames rather than performing the separation in configurational space needed to calculate $\Delta G_{ij, A}$ and $\Delta G_{ij, B}$.

\begin{figure}
    \centering
    \includegraphics[width=0.8\linewidth]{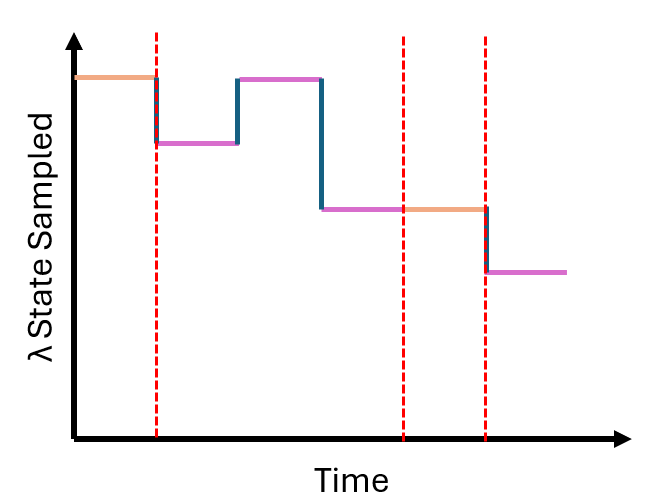}
    \begin{singlespace}
    \caption{This diagram shows how we can discretize trajectories based on whether they are sampling from conformational space A(orange) or B(pink). Over the course of a single iteration the system will perform Monte Carlo transitions between $\lambda$ states which is represented by movement along the y-axis. The system will also freely move in configurational space between space A and B shown by changes in vertical line color. In analysis we can separate the trajectory into frames which sample in configurational space A and B (dashed red lines). We concatenate the energies for all $\lambda$ states sampled in configurational space A and B and then perform separate MBAR analysis to obtain the free energy estimates.}
    \label{fig:sep_conf}
    \end{singlespace}
\end{figure}

\begin{figure}
    \centering
    \includegraphics[width=0.9\linewidth]{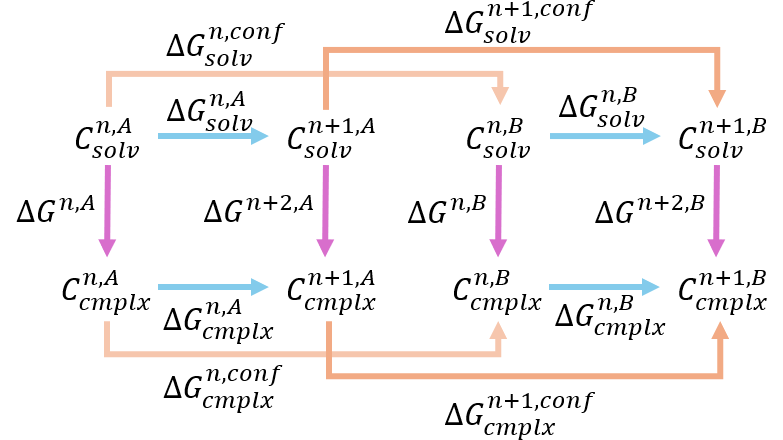}
    \caption{The separation of the free energy into conformational and alchemical free energy produces three distinct free energy cycles. One cycle follows the blue and pink arrows computing the relative free energy between C\textsubscript{n} to C\textsubscript{n+2}. The other two cycles follow the pink and either light or dark orange arrows computing the relative free energy for C\textsubscript{n} between pockets A and B and separately C\textsubscript{n+2} and pockets A or B.}
    \label{SI:cycle}
\end{figure} 

\begin{figure}
    \centering
    \includegraphics[width=0.8\linewidth]{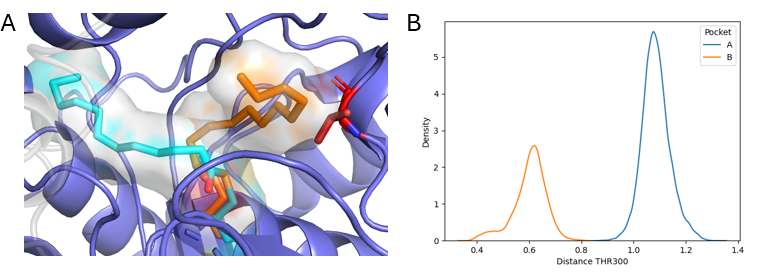}
    \caption{(A) We highlight the location of our reference residue THR300 (blue) position relative to the chain conformation in pocket A (orange) compared to pocket B (purple). (B) We plot the distance distributions in a set of trajectories with chain lengths ranging from 2 to 16 carbons and known to be in pocket A or B. The independent peaks demonstrate that this is an ideal metric for determining quantitatively defining the two pockets.}
    \label{SI:pocket_dist}
\end{figure}

\begin{figure}
    \centering
    \includegraphics[width=0.9\linewidth]{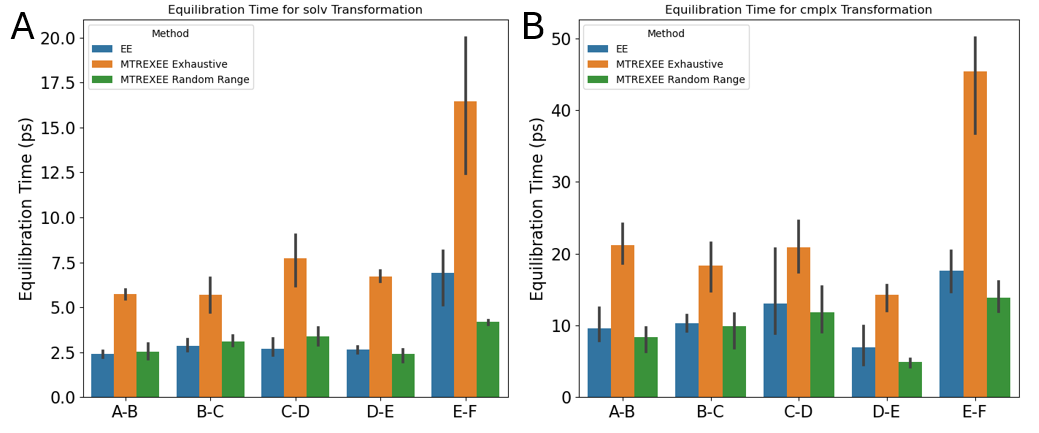}
    \caption{We compare the Wang-Landau weight equilibration times for all MUP1 small molecule ligand transformations performed both in solvent (A) and in the protein complex (B). The weight equilibration times for the MT-REXEE method with the Random Range equilibration method and no redundant end-states results in either statistically equivalent or lower equilibration times.}
    \label{SI:mup1_equil}
\end{figure}

\begin{figure}
    \centering
    \includegraphics[width=0.9\linewidth]{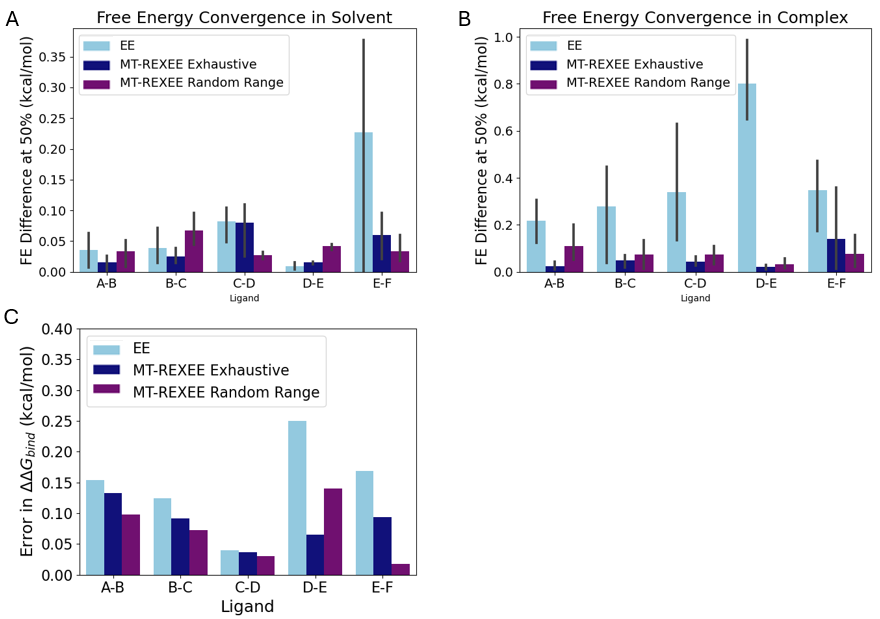}
    \caption{Here we use the absolute value of the difference in FE between using the first and last 50\% of the trajectory following Wang-Landau weight equilibration. If the estimates are closer together at 50\% we can conclude that the free energy estimates are converging more quickly which is likely a result of increased conformational sampling. (A) In the solvent simulations with the MUP1 ligands we see no real significant difference in the rate of convergence between methods. However in the protein complex simulations (B), we see a significant decrease in the difference between the forward and reverse estimates which indicates faster conformational sampling with MT-REXEE. (C) We also use the standard error between replicates as a measure of convergence for the free energy simulations which shows a significant decrease for the exhaustive method and an even more drastic decrease for MT-REXEE RR method.}
    \label{SI:mup1_converge}
\end{figure}

\begin{figure}
    \centering
    \includegraphics[width=0.9\linewidth]{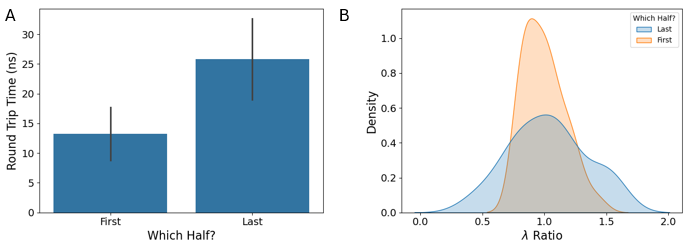}
    \caption{(A) The round trip time is the time it takes for a conformation to swap through all parallel simulations and return to the original $\lambda$ state. The round trip times are divided into the first and last half of the trajectories. We see a significant increase in the round trip time in the latter half of the simulation. (B) We compare the $\lambda$ ratio for all $\lambda$ states in each transformation in the first and second half of the trajectory. This is computed as $R = \frac{N_i}{N_{avg}}$ where $N_i$ is the number of frames sampled at a given $\lambda$ states, and $N_{avg}$ is the average frames sampled for all $\lambda$ states. In an ideal situation R = 1 because every state is sampled evenly. A value of R<1 means the state is under sampled and R>1 means the state is over-sampled. We observe a corresponding shift in the observed $\lambda$ ratio in the latter half of the simulation. This is possibly because the sampled conformational space is deviating significantly from the space sampled as the weights were equilibrating.}
    \label{SI:wl}
\end{figure}

\begin{figure}
    \centering
    \includegraphics[width=0.8\linewidth]{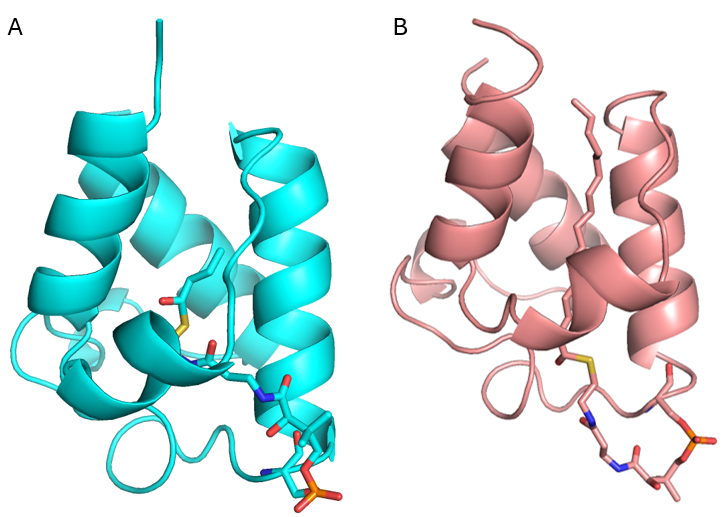}
    \caption{As the solvent reference state here we utilize the ACP with a sequestered acyl chain. We show centroid structures from the MT-REXEE simulations of the solvent state for C2-4 (A) and the C14-16 (B) states.}
    \label{SI:ACP_conf}
\end{figure}

\begin{figure}
    \centering
    \includegraphics[width=0.9\linewidth]{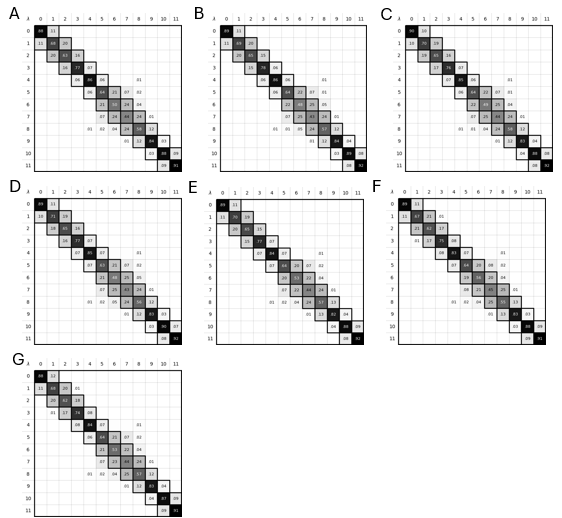}
    \caption{We measure the configurational overlap between $\lambda$ states in the ACP solvent simulations for transformations (A) A-B, (B) B-C, (C) C-D, (D) D-E, (E) E-F, (F) F-G, (G) G-H. We optimized the $\lambda$ state spacing in order to ensure off-diagonal values of at least 0.05.}
    \label{SI:ACP_solv_overlap}
\end{figure}

\begin{figure}
    \centering
    \includegraphics[width=0.9\linewidth]{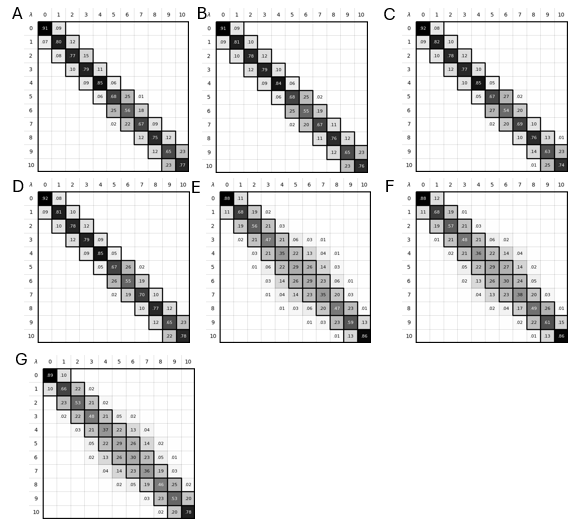}
    \caption{We measure the configurational overlap between $\lambda$ states in the FabB complex simulations for transformations (A) A-B, (B) B-C, (C) C-D, (D) D-E, (E) E-F, (F) F-G, (G) G-H. We optimized the $\lambda$ state spacing in order to ensure off-diagonal values of at least 0.05.}
    \label{SI:FabB_cmplx_overlap}
\end{figure}

\begin{figure}
    \centering
    \includegraphics[width=0.95\linewidth]{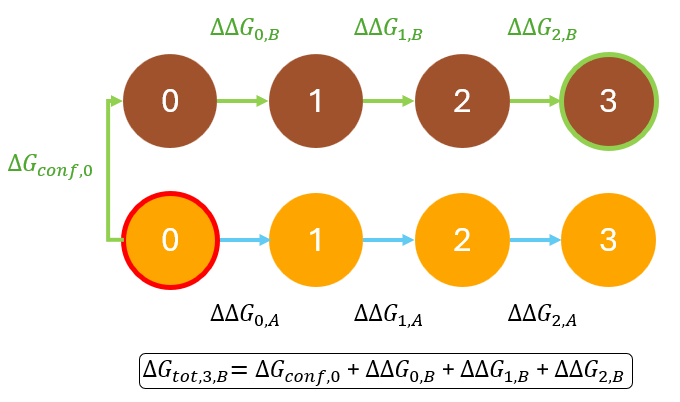}
    \caption{We present an example schematic for computing the cumulative FE difference to a single reference point as used in figure \ref{fig:FabB_FE}. In this example we compute the relative free energy difference between a C2 chain in pocket A and a C6 chain in pocket B. We add the configurational FE difference between pockets A and B for a C2 chain ($\Delta G_{conf, 0}$) to the alchemical free energy difference of performing the C2 to C4, C4 to C6, and C6 to C8 transformations in pocket B ($\Delta G_{0,B}$, $\Delta G_{1,B}$, $\Delta G_{2,B}$ respectively).}
    \label{SI:total_FE}
\end{figure}

\end{document}